\documentclass{IEEEtran}
\usepackage{graphicx}
\usepackage{subcaption}
\usepackage{amsmath}
\usepackage{amssymb} 

\begin{document}
\title{High-Bandwidth, Low-Computational Approach: Estimator-Based Control for  Hybrid Flying Capacitor Multilevel Converters Using Multi-Cost Gradient Descent and State Feedforward}

\author{Inhwi~Hwang, {inhwi@umich.edu}}

\markboth{Power Electronics and Control Report, 2024}%
{Shell \MakeLowercase{\textit{et al.}}: Bare Demo of IEEEtran.cls for IEEE Journals}

\maketitle

\begin{abstract}
This paper presents an estimator-based control framework for hybrid flying capacitor multilevel (FCML) converters, achieving high-bandwidth control and reduced computational complexity. Utilizing a hybrid estimation method that combines closed-loop and open-loop dynamics, the proposed approach enables accurate and fast flying capacitor voltage estimation without relying on isolated voltage sensors or high-cost computing hardware. The methodology employs multi-cost gradient descent and state feedforward algorithms, enhancing estimation performance while maintaining low computational overhead. A detailed analysis of stability, gain setting, and rank-deficiency issues is provided, ensuring robust operation across diverse converter levels and duty cycle conditions. Simulation results validate the effectiveness of the proposed estimator in achieving active voltage balancing and current control with 6-level AC-DC buck FCML, contributing to cost-effective solutions for FCML applications, such as data centers and electric aircraft.
\end{abstract}

\begin{IEEEkeywords}
flying capacitor multilevel converter (FCML), estimator-based control, active voltage balancing, state feedforward, multi-cost gradient descent method, hybrid estimatior, AC-DC buck conversion, datacenter power delivery.
\end{IEEEkeywords}

\IEEEpeerreviewmaketitle

\section{Background}
\IEEEPARstart{H}{ybrid} flying capacitor multilevel (FCML) converters are attracting interest for their power efficiency, power density, lightweight structure, and scalability \cite{254717,lai1995,radi2014,stillwell2016_1}.

\begin{figure}[t]
    \centering
    \includegraphics[width=1\linewidth]{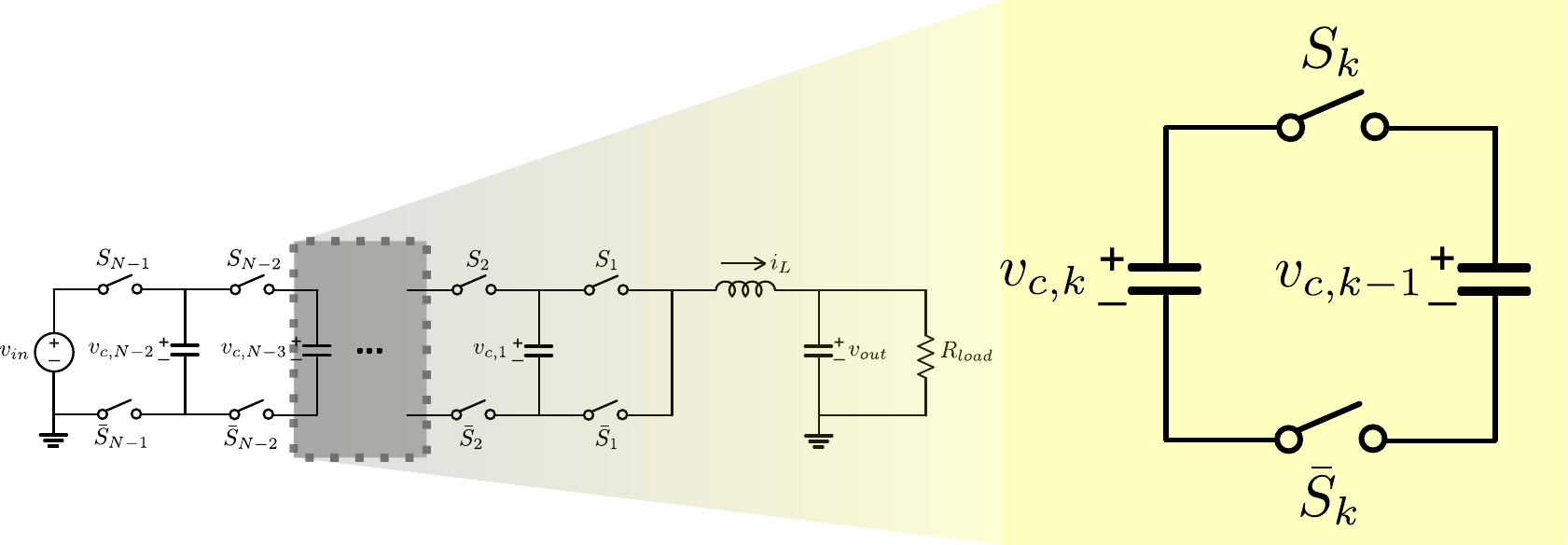}
    \caption{Single swiching cell of flying capacitor converter with adjacent flying capacitors and $k$-th switch pair. $v_{c,k}$ is the voltage of $k$-th flying capacitor voltage. $S_{k}$ and $\bar{S}_{k}$ are the switching states of $k$-th upper switch and lower switch, respectively.}
    \label{fig:single_cell}
\end{figure}

An $N$-level hybrid FCML converter employs $(N-2)$ flying capacitors to evenly distribute voltage stress across $(N-1)$ lower-voltage switches. As shown in Fig.~\ref{fig:single_cell}, the voltage across the $k$-th flying capacitor, where $k \in [1, N-2]$, is maintained at $\frac{k}{N-1}v_{in}$, with each switch experiencing a voltage stress of $\frac{1}{N-1}v_{in}$  \cite{lei2018}. The FCML topology also effectively spreads switching losses across multiple switches, enhancing thermal management. Additionally, power density is increased due to the reduced filter size, scaling by a factor of $(N-1)^2$ \cite{7846223}. Cascaded bootstrap gate drivers can further enhance the compactness of FCML hardware by removing the need for isolated DC-DC converters in gate drive circuits, contributing to reduced hardware complexity and design cost \cite{8911516}.

Utilizing the advantages of FCML, its applications are expanding across various fields. In spacecraft, FCML converters efficiently handle high voltage while maintaining a compact footprint, which is crucial for space-limited environments \cite{9487143,pallo2018}. Additionally, GaN-based eHEMT devices commonly used in FCML are radiation-hardened, ensuring reliable operation under high-radiation conditions encountered in space \cite{lidow2014}. For electric aircraft, FCML’s high power density supports lightweight designs and efficient space utilization, enhancing both efficiency and control performance \cite{10131430}. In data centers, FCML converters simplify the conventional two-stage step-up/down conversion process to single-stage step-down, reducing both system volume and complexity \cite{10613943,10221009,9591384}.

A challenge in hybrid FCML topology is ensuring \textbf{\textit{voltage balance across the flying capacitors}}. Each switch pair experiences voltage stress (${{v}_{stress,k}}$), as shown in Fig.~\ref{fig:single_cell}, determined by the voltage difference between adjacent flying capacitors as follows:
\begin{equation}
    {{v}_{stress,k}}={{v}_{c,k}}-{{v}_{c,k-1}}
\end{equation}
where $v_{0} = 0$. This voltage balance must be maintained in all situations, including start-up, closed-loop operation, and shut-down.

During start-up and shut-down, both the input and flying capacitors must charge or discharge evenly; otherwise, voltage imbalances may arise, increasing the stress on switching devices and risking overvoltage failure. To prevent this, each capacitor—including the input and flying capacitors—must charge in specific voltage ratios \cite{8371641, 10159129}.

\begin{figure*}[tbp]
    \centering
    \includegraphics[width=1\textwidth]{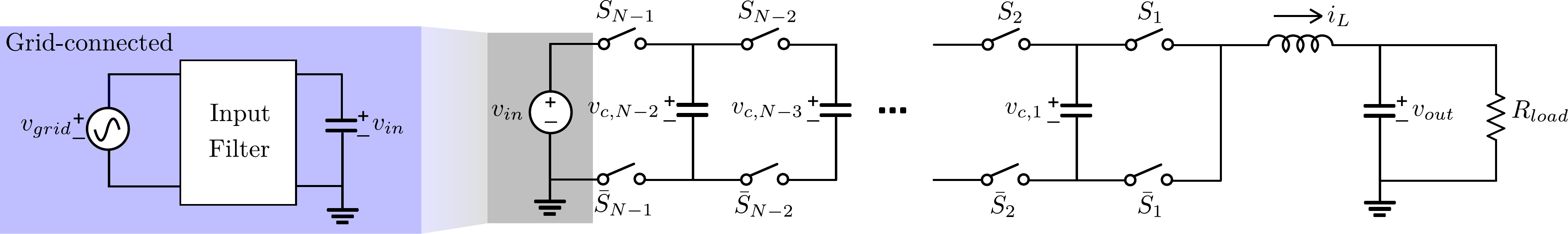}
    \caption{Circuit diagram of grid-connected buck-type hybrid FCML converter with input filter. $v_{grid}$, $v_{in}$, $i_L$, and $v_{out}$ are grid voltage, input capacitor voltage, inductor current, and output capacitor voltage, respectively.}
    \label{fig:buck_fcml}
\end{figure*}

For example, in an AC-DC buck FCML for data center power delivery, as illustrated in Fig.~\ref{fig:buck_fcml}, the start-up pre-charging process varies based on the output voltage condition. If the output voltage is connected to a 48V DC bus or pre-charged to 48V, a boosting algorithm (utilizing buck/boost duality) can charge the flying capacitors using the output energy. However, if the system must rely solely on AC power for initial charging, it is essential to limit inrush current through the input capacitor while ensuring adequate charging of the flying capacitors.

\begin{figure}[tbp]
    \centering
    \includegraphics[width=1\linewidth]{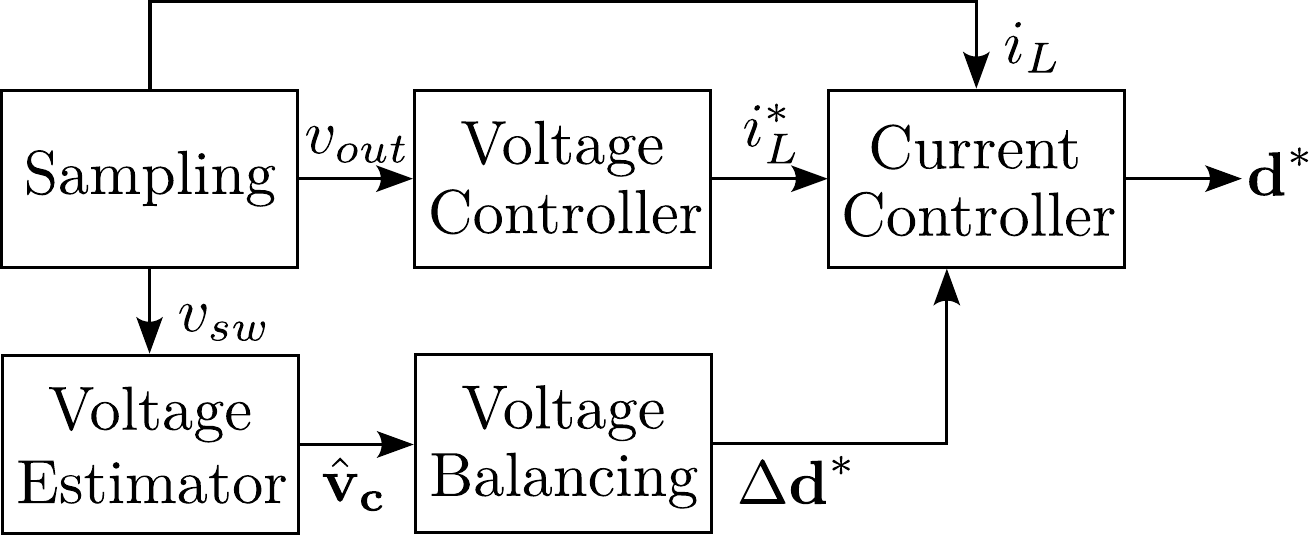}
    \caption{Block diagram of the estimator-based controller for hybrid FCML converter. $v_{sw}$ represents the pole voltage, while $\mathbf{\hat{v}_{c}}$ denotes the estimated flying capacitor voltage. $\mathbf{\Delta d^{*}}$ is the output of the voltage balancing controller, $\mathbf{d^{*}}$ is the duty cycle reference, and $i^*_L$ refers to the current control reference. The controller structure is hierarchically organized based on the principle of time-scale separation.}
    \label{fig:block_diagram}
\end{figure}

Additionally, voltage balancing is also essential during steady-state operation. For a DC input voltage that requires only maintaining DC flying capacitor voltage levels, passive balancing generally minimizes the control effort needed to sustain voltage balance. However, in grid-tied AC-DC buck converters, where $v_{in}$ oscillates at twice the line frequency, passive balancing alone cannot adequately maintain the correct flying capacitor voltage ratios, even under steady-state conditions \cite{xia2019}. The limited bandwidth of passive balancing increases the risk of overvoltage stress on switching devices. Consequently, a faster and more dynamic voltage balancing approach is necessary to ensure reliable operation and reduce the risk of device failure.

To achieve sufficient bandwidth, several methods for active balancing flying capacitor voltages have been introduced \cite{khazraei2012active, farivar2017capacitor, stillwell2019active, 10601504}. One promising approach uses closed-loop active balancing control and differential-mode voltage is utilized for feedforward term in current controller, achieving fast voltage balancing without impacting current control \cite{10601504}. This method offers higher bandwidth compared to conventional techniques. However, implementing the method \textbf{\textit{requires isolated voltage sensors}} for each of the $(N-2)$ flying capacitors due to the floating nature of each node’s voltage relative to ground. The use of these isolated voltage sensors increases hardware complexity and cost. 

To address these limitations, \textit{\textbf{estimator-based control}}  can be considered. The estimator-based control has become a widely adopted in power electronics field, such as motor control and grid-connected converters \cite{9999537,10509019}. By reducing the dependency on physical sensors, the estimator-based control offers cost-effectiveness and improved system reliability.

In literature \cite{9829954}, \textbf{\textit{flying capacitor voltage estimation}} method has been proposed. Although this method avoids the need for high-cost control units and complex implementation (e.g., FPGA) as seen in \cite{farivar2017capacitor, xia2019state}, it still presents certain challenges:

\begin{itemize}
    \item[1.] The method requires dual CPU operation within the MCU, resulting in additional communication overhead, as well as higher CPU and peripheral resource consumption.
    \item[2.] Precise sampling and rapid computations on one of the CPUs at rates of several hundred kHz are necessary, imposing a significant computational load.
    \item[3.] It relies solely on pole voltage data, overlooking other available information that could enhance the estimation of flying capacitor voltage.
    \item[4.] The literature has focused on implementing real-time estimation without offering mathematical proof of stability or clear guidelines for setting control gains.
    \item[5.] The literature has focused solely on estimation without exploring estimator-based control.
\end{itemize}

This paper addresses key research gaps by proposing an high-bandwidth, low-computation solution that operates with a single MCU CPU, without requiring additional peripheral resources. The method achieves high-bandwidth estimation with reduced sampling and control frequencies by utilizing given plant dynamics, duty cycle, and sampled inductor current information. This approach can enhance the versatility of estimator-based control for hybrid FCML converters, supporting a broad range of applications. The estimator-based control enables high-bandwidth operations such as current control and active voltage balancing, comparable to the performance achieved with sensor-based control.

Furthermore, this study includes a mathematical analysis from an optimization perspective, covering time-scale separation in estimator-based control, gradient descent, and estimator gain setting. Additionally, it examines the feasibility of achieving full-rank operation for each level of FCML under specific duty constraints.

The remainder of this paper is organized as follows. Chapter II discusses a hierarchical control structure based on time-scale separation principle. With generalized proportional-integral-resonant (PIR) controller for estimator-based control, controller design considerations are addressed based on application requirements. Chapter III introduces the proposed flying capacitor voltage estimator and a related sampling method and provides proofs of observability and stability, along with an analysis of the rank-deficiency problem and guidelines for gain settings. Chapter IV presents the simulation results that verify the effectiveness of the proposed method. Chapter V is the conclusion.

\begin{figure*}[t]
    \centering
    \includegraphics[width=\textwidth]{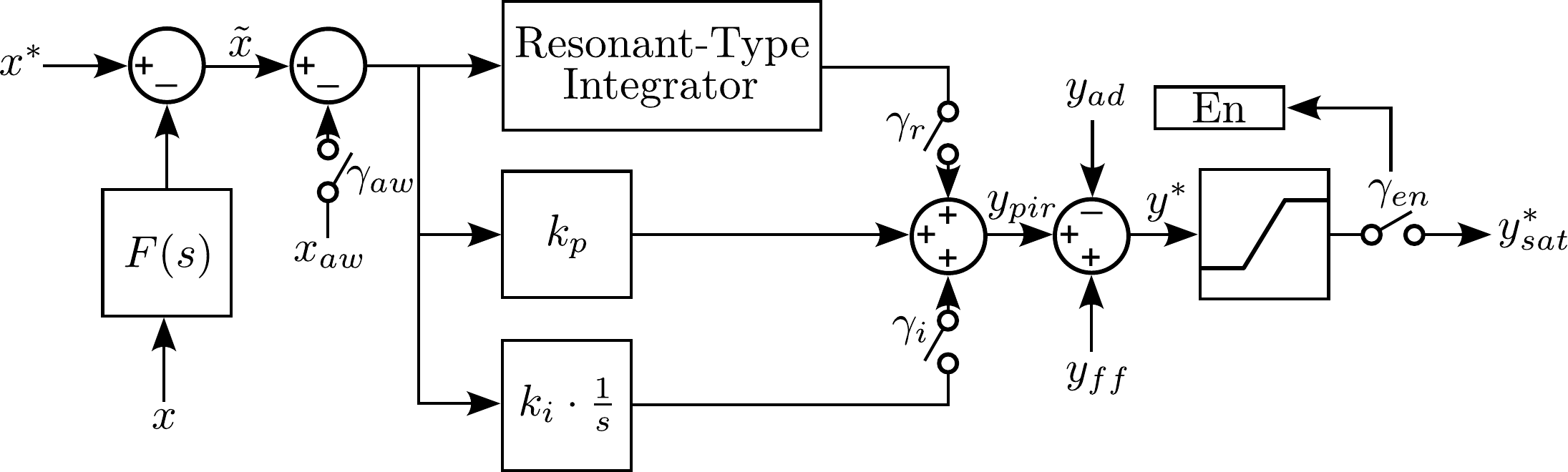}
    \caption{Block diagram of generalized proportional-integral-resonant (PIR) controller. $x$ is the target variable under control and $y$ is the output variable for controlling $x$. $\gamma$ is the variable to enable and disable some parts of the generalized PIR controller. $F(s)$ is the transfer function of the input filter.}
    \label{fig:controller}
\end{figure*}

\section{Estimator-Based Control: Hierarchical Controller Design and Considerations}
\subsection{Time-scale Separation}
The hybrid FCML’s multivariable control objectives, including the flying capacitor voltages, output voltage, and inductor current, present a challenging control problem due to the multiple control inputs and outputs in this multi-inpue multi-output (MIMO) system. The design problem can be simplified by organizing controllers in a cascaded loop, as shown in Fig.~\ref{fig:block_diagram}. This setup allows each control layer to be designed independently: 
\begin{itemize}
    \item The inner loop controller is assumed to have infinite bandwidth when designing the outer loop controller.
    \item The outer loop estimator/sampler is assumed to have infinite bandwidth when designing the inner loop controller.
\end{itemize}
These assumptions are applied to FCML controllers and estimator as follows:
\begin{equation}
\label{time_sep1}
    {{\tau }_{CC}}\ll {{\tau }_{VC}}
\end{equation}
\begin{equation}
\label{time_sep2}
    {{\tau }_{VE}}\ll {{\tau }_{VB}}
\end{equation}
Here, the settling time, $\tau$, is the inverse of the control bandwidth, $\omega$. The subscripts $CC$, $VC$, $VB$, and $VE$ represent the current controller, voltage controller, active balancing controller, and flying capacitor voltage estimator, respectively.

Meanwhile, in digital control systems, delays can arise that are not typically present in continuous-time systems. Control variables, such as inductor current, are sampled using a zero-order hold, and the pole voltage reference generated by the current controller introduces specific delays:
\begin{itemize}
    \item The reference is updated as a PWM comparator input in the next sampling period, introducing a one-sample delay.
    \item The average PWM voltage is applied halfway through the sampling period, resulting in a cumulative delay of 0.5 sampling periods, which directly impacts the controller's stability margin.
\end{itemize}
Therefore, the PWM pole voltage is effectively delayed by 1.5$\tau_s$. 
To mitigate instability caused by these delays, the sampling period should be much shorter than the controller’s settling time, as indicated by:
\begin{equation}
\label{time_sep3}
    {{\tau }_{s}}\ll {{\tau }_{CC}}
\end{equation}
where ${\tau }_{s}$ is the sampling period. According to the time-scale separation principle in \eqref{time_sep1}, \eqref{time_sep2}, and \eqref{time_sep3}, each controller’s bandwidth can be maximized while time-scale separation principle minimizes stability impacts between control layers while ensuring a fast response.

\begin{table*}[t]
\centering
\caption{Generalized PIR Controller Variables and Functions}
\resizebox{0.7\textwidth}{!}{%
\begin{tabular}{|c|c|l|}
\hline
\textbf{Symbol} & \textbf{Description} & \textbf{Explanation} \\
\hline
$x$ & Controller Input & Main input variable to the controller. \\
$x^{*}$ & Input Reference & Target or reference value for the input. \\
$\tilde{x}$ & Input Error & Difference between $x^{*}$ and $x$. \\
$x_{aw}$ & Anti-windup Input & Input used to limit integral windup effects. \\
\hline
$y$ & Controller Output & Main output variable from the controller. \\
$y^{*}$ & Output Reference & Target or reference value for the output. \\
$y_{pir}$ & PIR Controller Output & Output from the PIR controller block. \\
$y_{ff}$ & Feedforward Input & Direct feedforward input to the controller. \\
$y_{ad}$ & Active Damping Input & Input used for active damping control. \\
$y^{*}_{sat}$ & Final Output Reference & Saturated final output reference value. \\
\hline
$\gamma_{aw}$ & Anti-windup Switch & Enables/disables anti-windup function. \\
$\gamma_{r}$ & Resonant Integrator Switch & Enables/disables resonant integrators. \\
$\gamma_{i}$ & Integrator Switch & Enables/disables integrators. \\
$\gamma_{en}$ & Controller Enable Switch & Enables/disables the entire controller. \\
\hline
\end{tabular}%
}
\label{note}
\end{table*}

\subsection{Sampling}
To reduce the impact of switching ripple in sampled inductor current, the sampling frequency is typically synchronized with the PWM carrier, with sampling taking place at the peak or valley of the PWM carrier \cite{Sul2011}. For phase-shifted PWM (PSPWM), the sampling period (${\tau }_{s}$) is therefore aligned with the PWM carrier period (${\tau }_{sw}$) as follows:
\begin{equation}
\label{sampling}
{{\tau }_{s}}=\frac{{{\tau }_{sw}}}{2(N-1)}m_{s}
\end{equation}
where $m_{s}$ is a positive integer.

Since the effective switching frequency of the FCML converter with PSPWM typically reaches several hundred kHz \cite{stillwell2017}, and given the computational requirements per control cycle, the sampling and control frequencies are generally set between 10 and 40 kHz to ensure sufficient real-time processing capacity. The choice of sampling frequency is based on the computational load and the capabilities of the digital signal processor (DSP) in use; here, the TMS3202837xX CPU from Texas Instruments is considered.
\subsection{Generalized Proportional-Integral-Resonant Controller}
\indent The following subsections outline the design of the controller and estimator for the FCML, with each controller following a generalized PIR (Proportional-Integral-Resonant) framework shown in Fig.~\ref{fig:controller}. Each controller can be modified according to specific control objectives. 

In this framework, $x$ and $y$ represent the input and output variables of the controller, respectively. Here, $x^{*}$, $\tilde{x}$, and $x_{aw}$ stand for the input reference, input error, and anti-windup input, respectively, while $y^{*}$, $y_{pir}$, $y_{ff}$, $y_{ad}$, and $y^{*}_{sat}$ denote the output reference, PIR controller output, feedforward input, active damping input, and the final output reference. Control switches $\gamma_{aw}$, $\gamma_{r}$, $\gamma_{i}$, and $\gamma_{en}$ are used to enabling functions for anti-windup, resonant integrators, integrators, and the overall controller, respectively. For reference values, a superscript * is used throughout this paper. The following TABLE \ref{note} summarizes this information for clarity.

The controller gain is set to match the desired bandwidth by appropriately placing poles in the Laplace domain, based on the closed-loop transfer function of the plant and controller. Detailed formula-based gain settings for each controller are skipped in this paper.
\subsection{Output Voltage Control}
The control problem and the plant dynamic equation are:
\begin{equation}
x = {{v}_{out}}, \quad x^{*} = {{v}^{*}_{out}}, \quad {{y}^{*}_{sat}} \approx {{i}^{*}_{L}}
\end{equation}
\begin{equation}
{{C}_{out}} \frac{d{{v}_{out}}}{dt} = {{i}_{L}} - \frac{{{v}_{out}}}{{{R}_{load}}}
\end{equation}
, respectively. Here, $i_{L}$, $C_{out}$, and $R_{load}$ denotes the inductor current, output capacitance, and load resistance, respectively.

For \textit{DC/DC operation} of the FCML \cite{stillwell2019active}, an integrator in the controller is essential to eliminate steady-state error when using a DC reference. However, during scenarios such as initial charging or sudden load changes, large voltage errors may push the voltage controller’s output beyond the current limit, which clamps the current reference and reduces the voltage controller’s effective bandwidth.

Additionally, during current reference clamping, error accumulation in the integrator can lead to overshoot or undershoot in the output voltage, even after reaching the target voltage $v^{*}_{out}$. This, referred to as `integrator wind-up,' can occur in any controller with an integrator and output clamping. To address this, an anti-windup mechanism can be applied to mitigate the the error accumulation. With these considerations, the voltage controller is designed as follows:
\begin{equation}
\left[ {{\gamma}_{aw}}, {{\gamma}_{r}}, {{\gamma}_{i}}, {{\gamma}_{en}} \right] = \left[ 1, 0, 1, 1 \right]
\end{equation}
\begin{equation}
{{y}_{sat}}^{*} =
\begin{cases}
   {{y}^{*}} & \text{if } \left| {{y}^{*}} \right| \le {{i}_{L,\max }} \\
   \operatorname{sgn} \left( {{y}^{*}} \right) \cdot {{i}_{L,\max }} & \text{otherwise}
\end{cases}
\end{equation}
where $\operatorname{sgn}$ represents the sign function, and $i_{L,\max}$ is the maximum available inductor current.

For \textit{AC/DC boost operation} (e.g., power factor correction), the only difference from DC/DC is the presence of an AC power flow component at twice the line frequency \cite{7846223}. To control the DC output voltage, this AC power component can be filtered out by $F(s)$ ($F(0) = 1$, $F(j2n\omega_{g}) = 0$), or current controller can have multi-resonant controller ($\gamma_{r} = 1$) to eliminate the odd harmonic current component from voltage controller. 

For unity power factor operation, the current reference is multiplied by a unit sinusoidal waveform whose phase matches the grid phase using a phase-locked-loop (PLL). The output of controller with current limitation is as follows:
\begin{equation}
{{y}_{sat}}^{*} =
\begin{cases}
   {y}^{*} \cdot \sin(\hat{\theta}_{g}) & \text{if } {{y}^{*}} \le {{i}_{L,\max }} \\
   \operatorname{sgn}({y}^{*}) \cdot {{i}_{L,\max }} \cdot \sin(\hat{\theta}_{g}) & \text{otherwise}
\end{cases}
\end{equation}
where $\hat{\theta}_{g}$ is estimated grid phase from the PLL. $\omega_{g}$ denotes nominal angular frequency of grid.

For \textit{AC/DC buck operation} with an output inductor \cite{10221009,9591384}, two key points are noted: 
\begin{itemize}
    \item Power transfer between the grid and the FCML only occurs when the grid voltage magnitude surpasses the output voltage (e.g., 48 V in data center applications)
    \item The input voltage (grid voltage folded by rectifier) varies at twice the line frequency.
\end{itemize}
This results in both current and voltage containing DC and AC components with its harmonics. To manage these characteristics effectively, a proportional-resonant-integral (PIR) controller can be utilized as follows: \begin{equation}
\left[ {{\gamma}_{aw}}, {{\gamma}_{r}}, {{\gamma}_{i}} \right] = \left[ 1, 1, 1 \right]
\end{equation}
The controller's output with current limitation is defined as follows:
\begin{equation}
\label{current_reference}
{{y}_{sat}}^{*} =
\begin{cases} 
    {y}^{*} \cdot \psi(\hat{\theta}_{g}) + \phi(\hat{\theta}_{g}) & 
    \text{if } {{y}^{*}} \le {{i}_{L,\max }} \\
    \operatorname{sgn} \left( {y}^{*} \right) \cdot {{i}_{L,\max }} \cdot \psi(\hat{\theta}_{g}) 
    + \phi(\hat{\theta}_{g}) & 
    \text{otherwise}
\end{cases}
\end{equation}
Here, $\psi$ denotes the current reference waveform, which is synchronized with the grid voltage to ensure a high power factor. $\phi$ compensates for the effects of the input capacitor and filter on the grid current \cite{9591384}. It is crucial to set $i_{L,\max}$ with consideration for any amplitude increase resulting from $\phi$.

Meanwhile, when $|v_{grid}| < v_{out}$, the FCML converter is unable to draw power from the grid. In this case, no control or estimation is needed, and all stored energy in the flying capacitors and inductors remains constant, except for the output capacitor, which is gradually discharged by the load. The controller can be disabled to prevent unnecessary operation and error accumulation in integrators as follows:
\begin{equation}
\gamma_{en} = (|v_{grid}| > v_{out})    
\end{equation}

\subsection{Active Voltage Balancing}
The control problem for flying capacitor voltages is defined as follows:
\begin{equation}
\mathbf{x} = \mathbf{\hat{v}_{c}}, \quad \mathbf{x^{*}} = \frac{\mathbf{k}}{N-1}v_{in}, \quad \mathbf{{{y}^{*}_{sat}}} = \mathbf{\Delta d^{*}} 
\end{equation}
where ${{\mathbf{v}}_{\mathbf{c}}}={{\left[ \begin{matrix}
   {{v}_{c,1}} & {{v}_{c,2}} & \ldots  & {{v}_{c,N-2}}  \\
\end{matrix} \right]}^{\mathbf{T}}}$ denotes the flying capacitor voltage vector, $\mathbf{k} = [1 \; 2 \; \ldots \; N-2]^\mathbf{T}$ is the scaling vector, and $\mathbf{\Delta d} = [\Delta d_{1} \; \Delta d_{2} \; \ldots \; \Delta d_{N-2}]^\mathbf{T}$ represents the duty cycle differences. Each element $\Delta d_{k} = d_{k+1}-d_{k}$ for $k\in \left[ 1,N-2 \right]$, where $d_{k}$ is the PWM duty cycle of the $k$-th switch shown in Fig. \ref{fig:buck_fcml}. Here, the hat symbol ($\hat{x}$) denotes an estimated value.

All controllers adhere to the principle of time-scale separation; therefore by enforcing $\mathbf{v_{c}} \approx \mathbf{\hat{v}_{c}}$, the following plant equation for flying capacitor voltages can be considered:
\begin{equation}
\label{flyingcap}
{{\mathbf{C}}_{\mathbf{f}}}\frac{d{{\mathbf{v}}_{\mathbf{c}}}}{dt} = {{i}_{L}} \mathbf{\Delta S}
\end{equation}
where $\Delta \mathbf{S}=\left[ \begin{matrix}
   \Delta {{S}_{1}} & \Delta {{S}_{2}} & \cdots  & \Delta {{S}_{N-2}}  \\
\end{matrix} \right]^{\mathbf{T}}$, and $\Delta {{S}_{k}}={{S}_{k+1}}-{{S}_{k}}$ for $k\in \left[ 1,N-2 \right]$. The averaged plant equation over a sampling period ($\tau_{s}$) becomes:
\begin{equation}
  {{\mathbf{C}}_{\mathbf{f}}}\frac{d\left\langle \mathbf{v_{c}} \right\rangle}{dt} = \left\langle {{i}_{L}} \right\rangle \mathbf{\Delta d}
  \label{utilized_vc}
\end{equation}
This indicates that changes in flying capacitor voltages ($\mathbf{v_{c}}$) are influenced by both the duty cycle difference ($\mathbf{\Delta d}$) and the inductor current ($i_L$). Since $i_{L}$ is controlled by the output voltage controller, $\mathbf{\Delta d}$ remains the only variable available for controlling $\mathbf{v_{c}}$. Therefore, the active balancing controller can be implemented simply with a proportional controller as follows:
\begin{equation}
\left[ {{\gamma}_{aw}}, {{\gamma}_{r}}, {{\gamma}_{i}} \right] = \left[ 0, 0, 0 \right]
\end{equation}
\begin{equation}
\mathbf{{y}_{sat}}^{*} =
\begin{cases}
   \mathbf{{y}^{*}} & \text{if } \left| \mathbf{{y}^{*}} \right| \le \mathbf{{\Delta}d_{\max }} \\
   \operatorname{sgn} \left( \mathbf{{y}^{*}} \right) \cdot \mathbf{{\Delta}d_{\max }} & \text{otherwise}
\end{cases}
\label{lim_bal}
\end{equation}
It becomes necessary to limit $\mathbf{\Delta d}$ with $\mathbf{{\Delta}d_{\max }}$ when low inductor current results in a large $\mathbf{\Delta d}$. Excessive $\mathbf{\Delta d}$ can lead to a loss of inductor current control. $\mathbf{{\Delta}d_{\max }}$ prevents duty cycle saturation, thereby preserving stability in the current controller.

For \textit{AC/DC buck operation}, which requires high-bandwidth active voltage balancing, maintaining accurate voltage balance becomes increasingly critical as the input voltage ($v_{in}$) rises, helping to minimize stress on switching devices. 

In contrast, at lower $v_{in}$ levels, there is more tolerance for voltage imbalance, allowing minor control inaccuracies without major impact. Additionally, when $v_{in}$ is low, $d_{N-1}$ operates near its maximum duty cycle of 1, so $\mathbf{\Delta d}$ can potentially cause $d_{N-1}$ to reach saturation. This occurs even though active balancing is less critical than current control under these conditions. With these considerations, the controller can be disabled as follows:
\begin{equation}
{\gamma}_{en} = |v_{grid}| > m \cdot v_{out}
\end{equation}
where $m > 1$ provides a margin for disabling the active voltage balancing controller.

\subsection{Current Control}
The control objective and plant equation for inductor current are defined as follows:
\begin{equation}
x = i_{L}, \quad x^{*} = i^{*}_{L}, \quad {{y}^{*}_{sat}} = d^{*}_{N-1}
\end{equation}
\begin{equation}
L\frac{d{{i}_{L}}}{dt} = {{S}_{N-1}}{{v}_{in}} - {{\left( \mathbf{\Delta S} \right)}^{\mathbf{T}}}{{\mathbf{v}}_{\mathbf{c}}} - {{v}_{out}}
\end{equation}
respectively. The time-averaged plant equation over a sampling period ($\tau_s$) is:
\begin{equation}
L\frac{d{\left\langle{i}_{L}\right\rangle}}{dt} = {{d}_{N-1}}\left\langle {{v}_{in}} \right\rangle - {{\left( \mathbf{\Delta d} \right)}^{\mathbf{T}}} {\left\langle{\mathbf{v}}_{\mathbf{c}}\right\rangle} - \left\langle{{v}_{out}}\right\rangle
\end{equation}
To reduce disturbances from the output voltage ($v_{out}$) and differential term of flying capacitor voltages, the following feedforward term ($y_{ff}$) is applied:
\begin{equation}
{{y}_{ff}} = - {{\left( \mathbf{\Delta d} \right)}^{\mathbf{T}}}  \frac{{\mathbf{\hat{v}}}_{\mathbf{c}}}{v_{in}} - \frac{{v}_{out}}{v_{in}}
\end{equation}
For estimator-based control, estimated flying capacitor voltages ($\mathbf{\hat{v}_c}$) are used in the feedforward term, replacing $\mathbf{v_{c}}$ as shown in \cite{10601504}. A reduction in estimation bandwidth or an increase in estimation error may introduce disturbances, potentially degrading the current controller’s effective bandwidth. The disturbance effect on current control becomes more severe when inductor has lower inductance \cite{10509019}. Therefore, a fast and accurate estimator is required for estimator-based control.

For \textit{DC/DC boost operation}, an integral controller is required for eliminating the DC steady-state error of the current. The controller configuration is as follows:
\begin{equation}
\left[ {{\gamma}_{aw}}, {{\gamma}_{r}}, {{\gamma}_{i}}, {{\gamma}_{en}} \right] = \left[ 1, 0, 1, 1 \right]
\end{equation}

For \textit{AC/DC boost operation}, where the current is primarily a sinusoidal AC component, a resonant controller is preferable to achieve zero AC steady-state error:
\begin{equation}
\left[ {{\gamma}_{aw}}, {{\gamma}_{r}}, {{\gamma}_{i}}, {{\gamma}_{en}} \right] = \left[ 1, 1, 0, 1 \right]
\end{equation}

In \textit{AC/DC buck operation} for power factor correction, the inductor current $i_{L}$ may include both DC components and even harmonics of the line frequency, as noted in \eqref{current_reference}. This setup makes an integrator and a resonant integrator ideal choices for achieving unity closed-loop gain at specific frequencies. The current controller configuration for this case is:
\begin{equation}
\left[ {{\gamma}_{aw}}, {{\gamma}_{r}}, {{\gamma}_{i}}, {{\gamma}_{en}} \right] = \left[ 1, 1, 1, (|v_{grid}| > v_{out}) \right]
\end{equation}

For all cases, parameter variations, such as changes in inductor resistance, can affect the actual bandwidth of the current controller. To maintain precise control of the bandwidth, active damping is applied as follows:
\begin{equation}
y_{ad} = R_{a} i_{L}
\end{equation}

Additionally, nonlinearities in the controller, such as output limits, anti-windup mechanisms, and output scaling with specific waveforms, can introduce instability or create limit cycles with harmonic generation \cite{khalil2002nonlinear}. This consideration is important for the implementation of all controllers and estimators. To prevent these nonlinear effects in the current controller, a high-gain proportional controller can be a practical alternative:
\begin{equation}
\left[ {{\gamma}_{aw}}, {{\gamma}_{r}}, {{\gamma}_{i}}, {{\gamma}_{en}} \right] = \left[ 0, 0, 0, (|v_{grid}| > v_{out}) \right]
\end{equation}
This configuration prevents wind-up issues by avoiding use of integrators in the current controller. It does not achieve unity gain at DC and exhibits lower gain as frequency increases. However, the integrator in the output voltage controller compensates by ensuring zero steady-state error for output voltage control.

To accurately regulate the output current based on $y^{*}_{sat}$ from the voltage controller, the poles of the current controller should be placed for an overdamping response.

\section{Flying Capacitor Voltage Estimation}
\subsection{Considerations for Estimator Implementation}
In designing a state estimator, it is essential to ensure \textbf{\textit{observability}}. This involves evaluating the number of variables to be estimated and verifying that the given system matrix has full rank. For an $N$-level FCML converter, which has ($N-2$) flying capacitors, ($N-2$) independent equations are required to ensure a system matrix rank of ($N-2$).

\begin{figure}[t]
    \centering
    \includegraphics[width=1\linewidth]{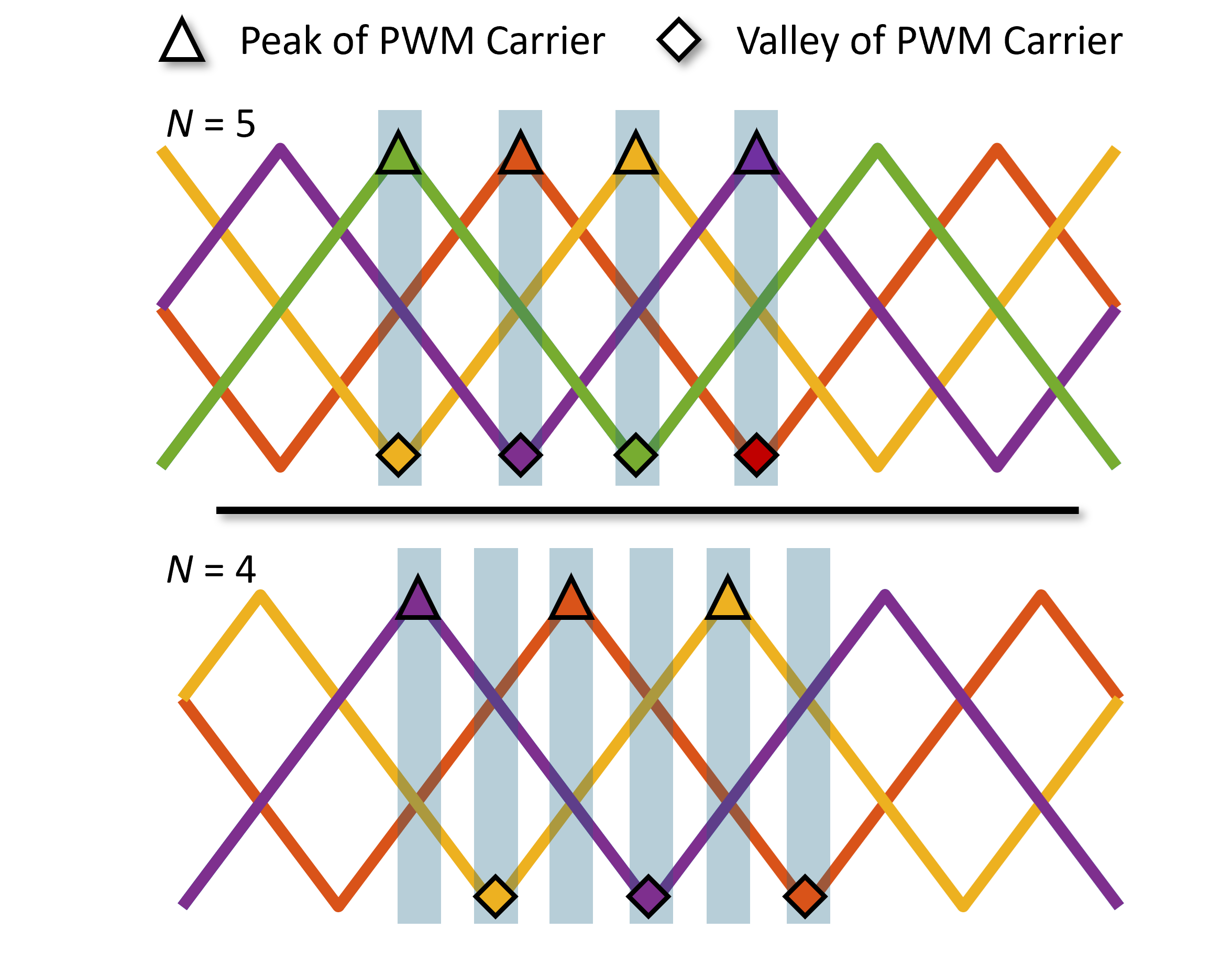}
    \caption{The figure of PSPWM carriers when $N = 5$ and $N = 4$. When \( N \) is an odd number, the peak of one PWM carrier coincides with the valley of another. Conversely, when \( N \) is an even number, no such overlap occurs, as the peaks and valleys are evenly distributed across the carriers.}
    \label{fig:odd_even}
\end{figure}

Secondly, once observability is confirmed with a full-rank, the implementation method must be considered in terms of \textbf{\textit{computational load and estimation performance}}, which has explicit trade-off. The simplest approach for estimating $\mathbf{v_c}$ is utilizing matrix inversion. However, as the number of levels in the FCML increases, the size of the state-space matrix grows significantly, containing at least $(N-2)^2$ elements. \\
\indent Common matrix inversion algorithms have theoretical complexities ranging from $O(n^{2.81})$ to $O(n^3)$ \cite{COSME2018125,6710599}. For example, with $N=8$, where $n=8-2=6$, matrix inversion requires between approximately $6^{2.81} \approx 154$ and $6^3 = 216$ operations. This rapid increase in computational load imposes a significant burden on the CPU of MCU. In industry, where low-cost CPUs are widely preferred, such processing demands are undesirable, as they would require a more complex and costly processing unit. This challenge highlights the need for computationally efficient estimation methods.\\
\indent In \cite{9829954}, a method with a theoretical complexity of $O(n)$ per sampling period was proposed to address the high computational load typically associated with matrix inversion. This real-time estimation technique offers an advantage by distributing computational complexity across multiple sampling instances rather than requiring full-rank conditions at each individual sample. As a result, it eliminates the need for matrix inversion while maintaining proper accuracy and bandwidth. Moreover, a balanced workload can be achieved by spreading calculations across control instances, making the computation easier to manage. However, the implementations of the method still rely on a high sampling rate and additional peripheral communication.\\
\indent Finally, a well-designed estimator must \textbf{\textit{utilize all available information}} to maximize the estimation performance. Fully utilizing the information improves the figure of merit for estimation, the superior balances can be found on the trade-off between the computational load and fast/accurate estimation.

In the proposed estimation method, the sampled current ($i_L$), duty cycle reference ($\mathbf{d^{*}}$), and flying capacitor voltage ($\mathbf{v_{c}}$) dynamics in (\ref{utilized_vc}) are utilized as new information. By incorporating this information, the estimation performance can be significantly enhanced, even with lower sampling and control frequency.

\subsection{Disjoint Sampling: Extracting Full-Rank System Equations}
\begin{figure*}[t]
    \centering
    \includegraphics[width=1\textwidth]{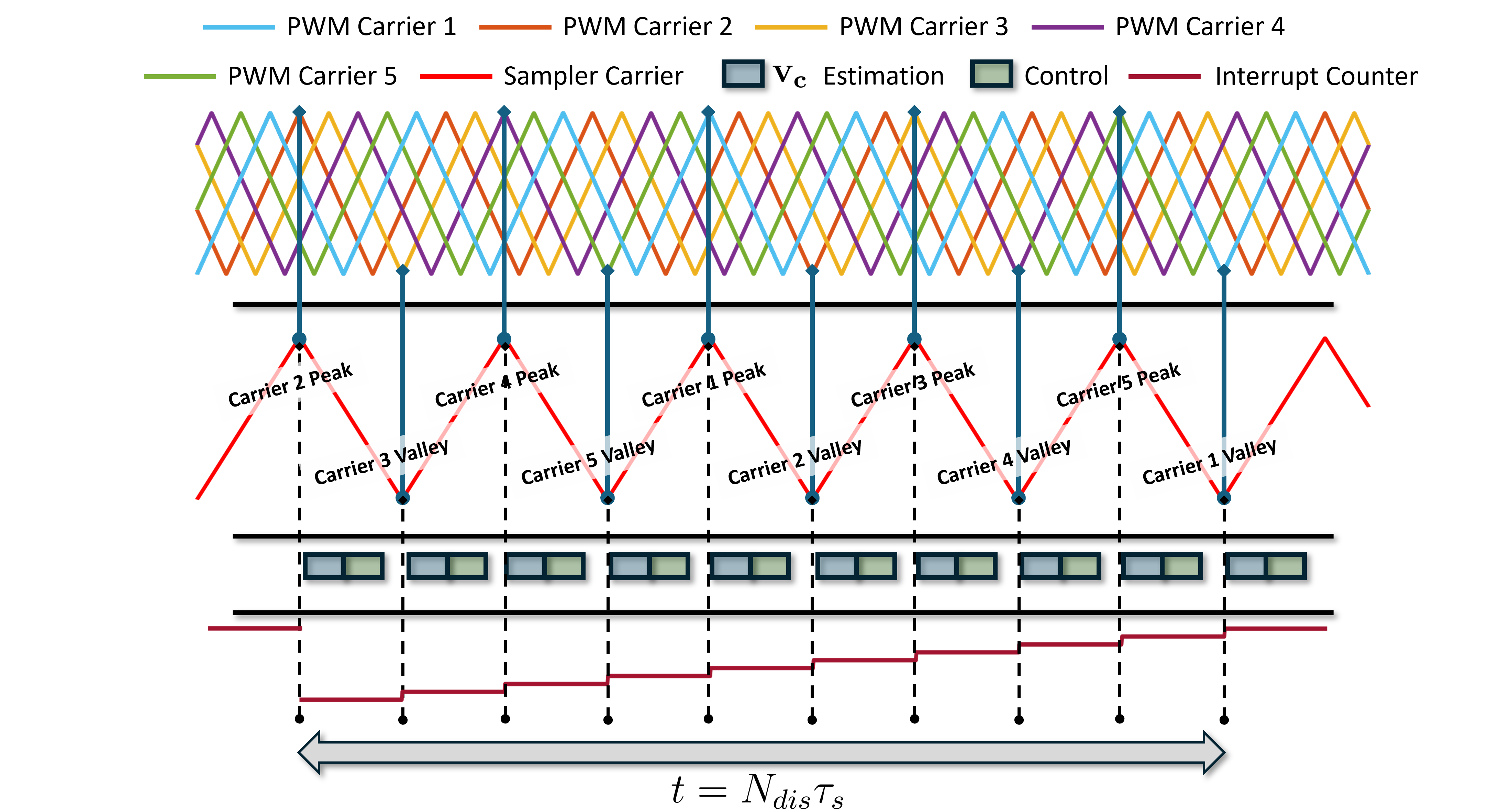}
    \caption{Disjoing sampling with \( m_s = 7 \) for 6-level FCML converter with 5 PSPWM carriers where \( N_{dis} = 10 \). Disjoint sampling ensures that all sampling points coincide with all the peaks and valleys of the PSPWM carriers. To determine whether a sampling point corresponds to a peak or valley of a specific carrier, a PWM interrupt counter is used.
}
    \label{fig:carriers}
\end{figure*}
\indent The pole voltage, a linear function of the flying capacitor voltages, is measured and used in the proposed estimator. A non-isolated voltage sensor can be utilized for this measurement. The pole voltage is sampled at the peak and valley of the $(N-1)$ PSPWM carriers, synchronously sampling voltage and current signals for control. At each sampling instance, estimator calculations are performed along with control operations, enabling estimation and control integration in a single control loop.

As shown in Fig.~\ref{fig:odd_even}, for an odd-level FCML, $(N-1)$ different equations can be obtained during peak and valley sampling for the same duty reference under PSPWM, while for an even-level FCML, $2(N-1)$ different equations can be obtained. The proposed disjoint sampling method shown in Fig.~\ref{fig:carriers} uses a sampler carrier synchronized with PSPWM carriers, operating at a frequency of $f_s$ which is much lower than effective switching frequency $(N-1)f_{sw}$. The number of different sampling instants ($N_{dis}$) is defined as follows:
\begin{equation}
\label{N_dis}
N_{dis} = 
\begin{cases}
2(N-1), & \text{if } N \text{ is even}, \\
N-1, & \text{if } N \text{ is odd}.
\end{cases}
\end{equation}
Sampling is triggered when the sampler carrier reaches its valley, incrementing the interrupt counter. To utilize disjoing sampling, $m_s$ in \eqref{sampling} is selected based on the following conditions:
\begin{equation}
N_s \in 
\begin{cases}
\mathbb{Z}^{+} \, \big| \, \gcd(N_s, 2(N-1)) = 1, & \text{if } N \text{ is even}, \\
\mathbb{Z}^{+} \, \big| \, \gcd(N_s, N-1) = 1, & \text{if } N \text{ is odd}.
\end{cases}
\end{equation}
\begin{equation}
m_s = 
\begin{cases}
N_{s}, & \text{if } N \text{ is even}, \\
2N_{s}, & \text{if } N \text{ is odd}.
\end{cases}
\end{equation}
The sampling frequency setting differs between odd-level and even-level FCML converters. This is because, as illustrated in Fig. \ref{fig:odd_even}, the even number of PSPWM carriers for odd-level FCML may result in one carrier's peak coinciding with another's valley. 

Meanwhile, the proposed method will utilize a lower sampling rate compared to \cite{9829954}, which makes the estimator’s closed-loop bandwidth and accuracy degraded. While the proposed method simplifies sampling implementation, it may not fully support high-bandwidth controllers based on time-scale separation. The reduced bandwidth due to lowered sampling frequency will be highly improved through state-feedforward, which will be introduced in the following chapter.
\subsection{Flying Capacitor Voltage Estimator}
\subsubsection{Multi-Cost Gradient Descent Method for Closed-Loop Estimation} From an optimization perspective, the convex optimization problem for estimating flying capacitor voltages can be formulated as follows:

\begin{equation}
\label{opt}
\underset{{{{\mathbf{\hat{v}}}}_{\mathbf{cl}}}\in {{\mathbb{R}}^{N-2}}}{\mathop{\min }}\,{{\left( {{{\mathbf{\hat{v}}}}_{\mathbf{cl}}}-{{\mathbf{v}}_{\mathbf{c}}} \right)}^{\mathbf{T}}}\mathbf{Q}\left( {{{\mathbf{\hat{v}}}}_{\mathbf{cl}}}-{{\mathbf{v}}_{\mathbf{c}}} \right),
\end{equation}
where $\mathbf{Q} > 0$, and ${\mathbf{\hat{v}}}_{\mathbf{cl}}$ denotes the estimated value of $\mathbf{{v}_{c}}$ of closed-loop estimator. Here, `$\mathbf{Q}>0$' means matrix $\mathbf{Q}$ is positive definite. The optimal value is 0, and the optimal solution is ${{\mathbf{\hat{v}}}_{\mathbf{cl}}}={{\mathbf{v}}_{\mathbf{c}}}$.
\begin{figure*}[t]
    \centering
    \includegraphics[width=0.8\textwidth]{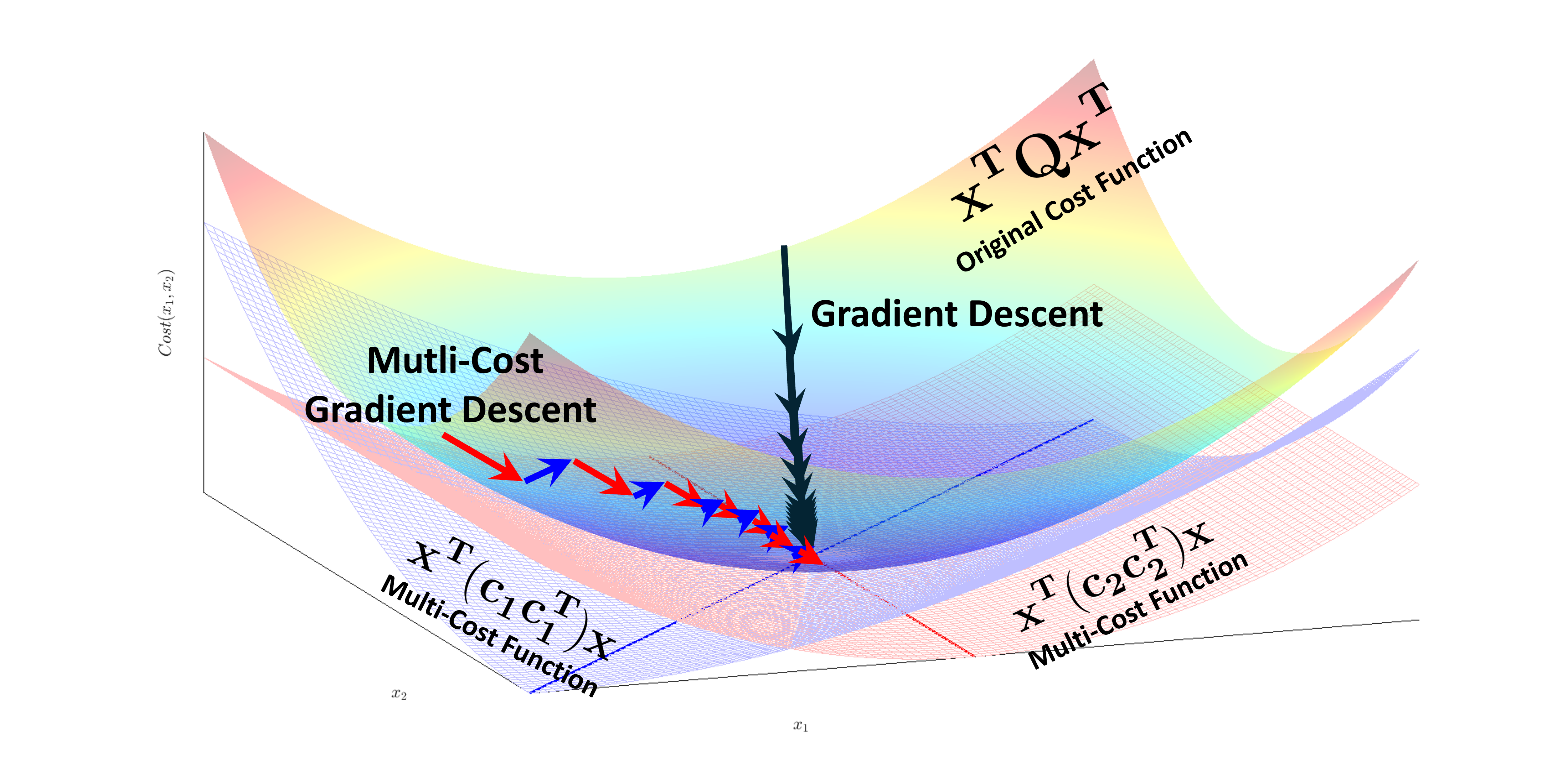}
    \caption{A graph showing the original cost function of convex optimization and multi-cost functions. The figure illustrates an example of convex optimization with two variables. Multi-cost matrices are all single rank. The optimal solution of the original cost function is shown to coincide with the intersection of the optimal solutions of the multi-cost functions. The gradient of each multi-cost function is constrained along a fixed direction vector, which is the eigenvector of the multi-cost matrix. 
}
    \label{fig:Grad}
\end{figure*}

Instead of solving the original optimization problem defined in (\ref{opt}), the proposed method proposes a multi-cost function approach shown in Fig.~\ref{fig:Grad}, defined as follows:
\begin{equation}
\underset{{{{\mathbf{\hat{v}}}}_{\mathbf{cl}}}\in {{\mathbb{R}}^{N-2}}}{\mathop{\min }}\,\sum_{l=1}^{{{N}_{dis}}}{{\left( {{{\mathbf{\hat{v}}}}_{\mathbf{cl}}}-{{\mathbf{v}}_{\mathbf{c}}} \right)}^{\mathbf{T}}}\left( {{\mathbf{c}}_{l}}{{\mathbf{c}}_{l}}^{\mathbf{T}} \right)\left( {{{\mathbf{\hat{v}}}}_{\mathbf{cl}}}-{{\mathbf{v}}_{\mathbf{c}}} \right), \quad l \in [1, ..., N_{dis}]
\label{mult_opt}
\end{equation}
where ${{\mathbf{c}}_{l}}{{\mathbf{c}}_{l}}^{\mathbf{T}} > 0$ ensures that each cost function is convex. The gradient descent method guarantees convergence to the optimal value, 0. The optimal condition is expressed as:

\begin{equation}
{{\mathbf{c}}_{l}}^{\mathbf{T}}\left( {{{\mathbf{\hat{v}}}}_{\mathbf{cl}}}-{{\mathbf{v}}_{\mathbf{c}}} \right)=0
\end{equation}
The intersection of solution is ${{\mathbf{\hat{v}}}_{\mathbf{cl}}}={{\mathbf{v}}_{\mathbf{c}}}$ and the optimal solution is 0, which is same as the original optimization problem in (\ref{opt}), only if the following matrix is invertible (observability requirement):

\begin{equation}
{{\left[ \begin{matrix}
   {{\mathbf{c}}_{1}} & {{\mathbf{c}}_{2}} & \ldots  & {{\mathbf{c}}_{{{N}_{dis}}}}  
\end{matrix} \right]}^{\mathbf{T}}}.
\label{invertible}
\end{equation}
This requires that:
\begin{equation}
\operatorname{span}\left\{ \mathbf{c}_{1}^{\mathbf{T}}, \mathbf{c}_{2}^{\mathbf{T}}, \ldots, \mathbf{c}_{{{N}_{dis}}}^{\mathbf{T}} \right\} = {{\mathbb{R}}^{N-2}}.
\label{span}
\end{equation}

For the proposed real-time estimation of flying capacitor voltages, the multi-cost gradient vectors $\mathbf{c}_{l} \in \left\{ \mathbf{c}_{1}, \mathbf{c}_{2}, \ldots, \mathbf{c}_{{{N}_{dis}}} \right\}$ are utilized. The closed-loop update function, based on the gradient descent method in the discrete-time domain ($t = n\tau{s}$), is expressed as follows:
\begin{equation}
\label{opt_upd}
\begin{aligned}
  & {{{\mathbf{\hat{v}}}}_{\mathbf{cl}}}\left[ n \right]\\
  &
  ={{{\mathbf{\hat{v}}}}_{\mathbf{cl}}}\left[ n-1 \right]-\mathbf{c}\left[ n \right]\cdot \alpha \mathbf{c}{{\left[ n \right]}^{\mathbf{T}}}\left( {{{\mathbf{\hat{v}}}}_{\mathbf{cl}}}\left[ n-1 \right]-{{\mathbf{v}}_{\mathbf{c}}}\left[ n \right] \right) \\ 
 & =\left( \mathbf{I}-\alpha \mathbf{c}\left[ n \right]\mathbf{c}{{\left[ n \right]}^{\mathbf{T}}} \right){{{\mathbf{\hat{v}}}}_{\mathbf{cl}}}\left[ n-1 \right]+\alpha \mathbf{c}\left[ n \right]\mathbf{c}{{\left[ n \right]}^{\mathbf{T}}}{{\mathbf{v}}_{\mathbf{cl}}}\left[ n \right] \\ 
\end{aligned}
\end{equation}
where $\mathbf{I}$ and $\alpha$ denotes the identity matrix and feedback gain (learning rate). In \eqref{opt_upd}, utilizing the pole voltage equation:
\begin{equation}
\label{pole_voltage}
v_{sw} = -{{\left( \Delta \mathbf{S} \right)}^{\mathbf{T}}}{{\mathbf{v}}_{\mathbf{c}}} + S_{N-1}v_{in},
\end{equation}
for setting the gradient vector $\mathbf{c}[n] = \Delta \mathbf{S}[n]$, the update function becomes:
\begin{equation}
\begin{aligned}
{{\mathbf{\hat{v}}}_{\mathbf{cl}}}[n] = & \left( \mathbf{I} - \alpha \mathbf{\Delta S}[n] \mathbf{\Delta S}^{\mathbf{T}}[n] \right) {{\mathbf{\hat{v}}}_{\mathbf{cl}}}[n-1] \\
& + \alpha \left( S_{N-1}v_{in} - v_{sw} \right) \mathbf{\Delta S}[n].
\end{aligned}
\end{equation} where $\alpha$ is learning rate, which is feedback gain.
This update function links the flying capacitor voltage estimation to pole voltage ($v_{sw}$) and input voltage ($v_{in}$). By utilizing the switching state vector ($\mathbf{\Delta S}[n]$), the gradient descent method can be effectively adapted to the physical characteristics of the FCML converter.

\begin{figure*}[tp]
    \centering
    \includegraphics[width=0.8\textwidth]{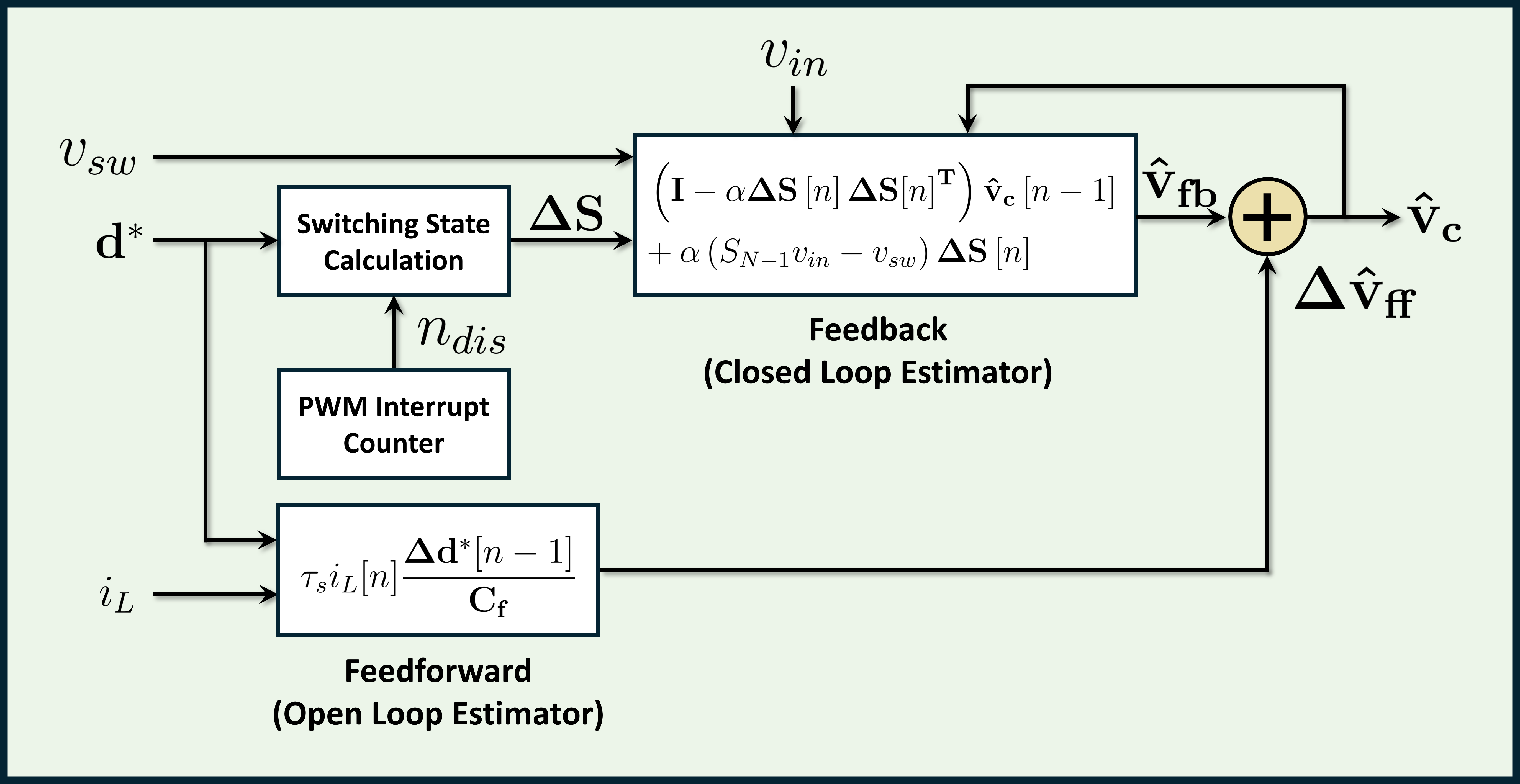}
    \caption{Block diagram of hybrid flying capacitor voltage estimator.}
    \label{fig:VE}
\end{figure*}
\begin{figure}[t]
    \centering
    \includegraphics[width=0.65\linewidth]{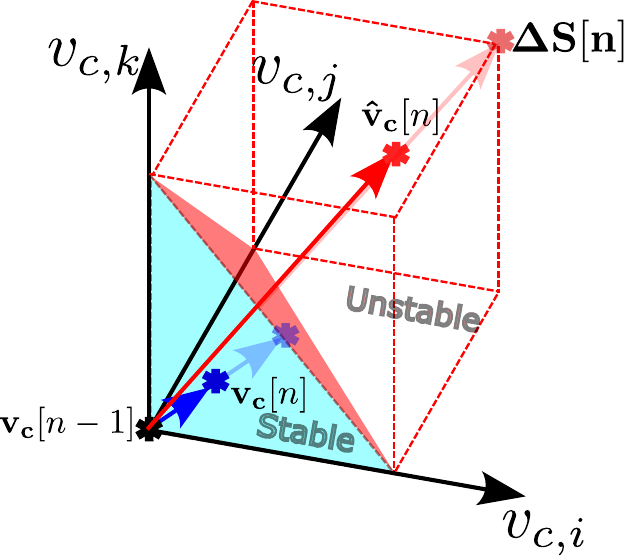}
    \caption{A figure illustrating closed-loop estimation of flying capacitor voltage where $v_{c,i},v_{c,j},v_{c,k}$ is the component of $\mathbf{v_{c}}$ with non-zero $\Delta S_{i}[n]$, $\Delta S_{j}[n]$, and $\Delta S_{k}[n]$.  The update vector (\( \mathbf{\hat{v}_c}[n] \) - \( \mathbf{\hat{v}_c}[n-1]\)) for the next prediction is the projection of the error vector (\( \mathbf{v_c}[n]\)-\( \mathbf{\hat{v}_c}[n-1]\)) onto $\mathbf{\Delta S}[n]$, scaled by \( \alpha \). The update vector is parallel to \( \mathbf{\Delta S}[n] \).
}
    \label{alpha_fig}
\end{figure}

\subsubsection{State-Feedforward for Open-Loop Estimation}
To enhance the dynamic response of the closed-loop estimator, an open-loop estimator is utilized for a state-feedforward.\\
\indent According to dynamics of flying capacitor voltage ($\mathbf{v_{c}}$) in (\ref{flyingcap}), the flying capacitors are charged and discharged by inductor current ($i_{L}$) based on the switching state ($\mathbf{\Delta S}$).
Over one sampling period, the change in charge of each flying capacitor is determined by the product of $\mathbf{\Delta d}[n]$ and $i_{L}[n]$. $\mathbf{\Delta d}[n]$ can be estimated by the previous information of duty reference $\mathbf{\Delta d^*}[n-1]$ with deadtime, therefore the flying capacitor voltage can be estimated in open-loop as follows:
\begin{equation}
\mathbf{\hat{v}_{ol}}\left[ n\right] = \mathbf{\hat{v}_{ol}}\left[ n-1\right] + \tau_{s} i_{L}[n] \frac{\mathbf{\Delta d^{*}}[n-1]}{\mathbf{C_{f}}}
\end{equation} where ${\mathbf{\hat{v}}}_{\mathbf{ol}}$ denotes the estimated value of $\mathbf{{v}_{c}}$ of open-loop estimator. 

\subsection{Hybrid Estimator}
By integrating the closed-loop and open-loop estimators, the final update function is derived as follows:
\begin{equation}
\label{updatefun}
\mathbf{\hat{v_{c}}}[n] = \mathbf{\hat{v}_{fb}}[n] + \mathbf{\Delta \hat{v}_{ff}}[n]
\end{equation}
\begin{equation}
\begin{aligned}
\mathbf{\hat{v}_{fb}}[n] &= \left( \mathbf{I} - \alpha  \mathbf{\Delta S}\left[ n \right] \mathbf{\Delta} \mathbf{S}{{\left[ n \right]}^{\mathbf{T}}} \right) \mathbf{\hat{v}_{c}}\left[ n-1 \right] \\
& \quad + \alpha\left( {{S}_{N-1}}{{v}_{in}} - {{v}_{sw}} \right) \mathbf{\Delta S}\left[ n \right]
\end{aligned}
\end{equation}
\begin{equation}
\mathbf{\Delta \hat{v}_{ff}}\left[ n\right] = \tau_{s} i_{L}[n] \frac{\mathbf{\Delta d^{*}}[n-1]}{\mathbf{C_{f}}}
\end{equation}
where $\mathbf{\hat{v}_{c}}$, $\mathbf{\hat{v}_{fb}}$, and $\mathbf{\hat{v}_{ff}}$ are the estimated flying capacitor voltage, the feedback and feedforward terms of hybrid estimator, respectively. The block diagram of the hybrid estimator is depicted in Fig.~\ref{fig:VE}.

The feedback term ensures convergence to actual value of the estimated values under steady-state conditions while guaranteeing stability of the estimator. The feedforward term compensates the limited dynamic response of the closed-loop estimator caused by lower sampling and control frequency by utilizing the fast dynamic of the open-loop estimation. 

The hybrid estimator combines the strengths of both methods, enabling rapid tracking of flying capacitor voltage changes with open-loop estimation, while ensuring stability and convergence with closed-loop estimation. This approach allows for achieving high-bandwidth performance even with low sampling and control rates, making high-bandwidth estimator-based control feasible with low-cost MCU. 

The mathematical analysis of the proposed hybrid estimator, including its stability and proper gain setting, will be addressed in detail in the following chapter.

\subsection{Stability Analysis}
\eqref{updatefun} can be expressed in different way to analyze the stability in discrete-time domain as follows:
\begin{equation}
\label{system_matrix}
\begin{aligned}
\mathbf{\hat{v}_{c}}\left[ n \right] &= \left( \mathbf{I} - \alpha \mathbf{\Delta S}\left[ n \right] \Delta \mathbf{S}{{\left[ n \right]}^{\mathbf{T}}} \right) \mathbf{\hat{v}_{c}}\left[ n-1 \right] \\
& + \alpha \mathbf{\Delta S}\left[ n \right] \Delta \mathbf{S}{{\left[ n \right]}^{\mathbf{T}}} \mathbf{v_{c}}\left[ n \right] \\
& + {{\mathbf{\Delta \hat{v}}}_{\mathbf{ff}}}\left[ n \right]
\end{aligned}
\end{equation}
Here, stability of the estimator is determined by the system matrix's eigenvalues. The system matrix is
\begin{equation}
\mathbf{P}\left[ n \right]=\frac{\partial {{{\mathbf{\hat{v}}}}_{\mathbf{c}}}\left[ n \right]}{\partial {{{\mathbf{\hat{v}}}}_{\mathbf{c}}}\left[ n-1 \right]}=\mathbf{I}-\mathbf{\alpha} \mathbf{\Delta }{\mathbf{S}}[n]\mathbf{\Delta }{\mathbf{S}}[n]^{\mathbf{T}} 
\label{commutative}
\end{equation}

Here, both $\mathbf{I}$ and $\mathbf{\Delta S \Delta S^{T}}$ are commutative and simultaneously diagonalizable, so the eigenvalues of system matrix are linear combinations of each matrix's eigenvalues. Each non-zero row vector of $\mathbf{\Delta S \Delta S^{T}}$ is all paralleled each other and have a same magnitude with $\mathbf{\Delta S^{T} \Delta S}$ and the rank of $\mathbf{\Delta S \Delta S^{T}}$ is 1. Therefore, the eigenvalues of $\mathbf{\Delta S \Delta S^{T}}$ are 0 and $\mathbf{\Delta S^{T} \Delta S }$. As a result, the eigenvalues of system matrix are
\begin{equation}
\lambda_{eig}[n] \in \{1-\alpha\mathbf{\Delta S}[n]^\mathbf{T} \mathbf{\Delta S}[n], 1\}
\end{equation}
To make sure the stability of the estimator, all eigenvalues should be in a range between -1 and 1, in other words,
$|\lambda_{eig}[n]|<1$, therefore the following inequation should be met:
\begin{equation}
0<\alpha\mathbf{\Delta S}[n]\mathbf{^{T}}\mathbf{ \Delta S}[n]<2
\label{eig_range}
\end{equation}

\begin{figure*}[tp]
    \centering
    
    \begin{subfigure}[t]{0.7\linewidth} 
        \centering
        \includegraphics[width=\linewidth]{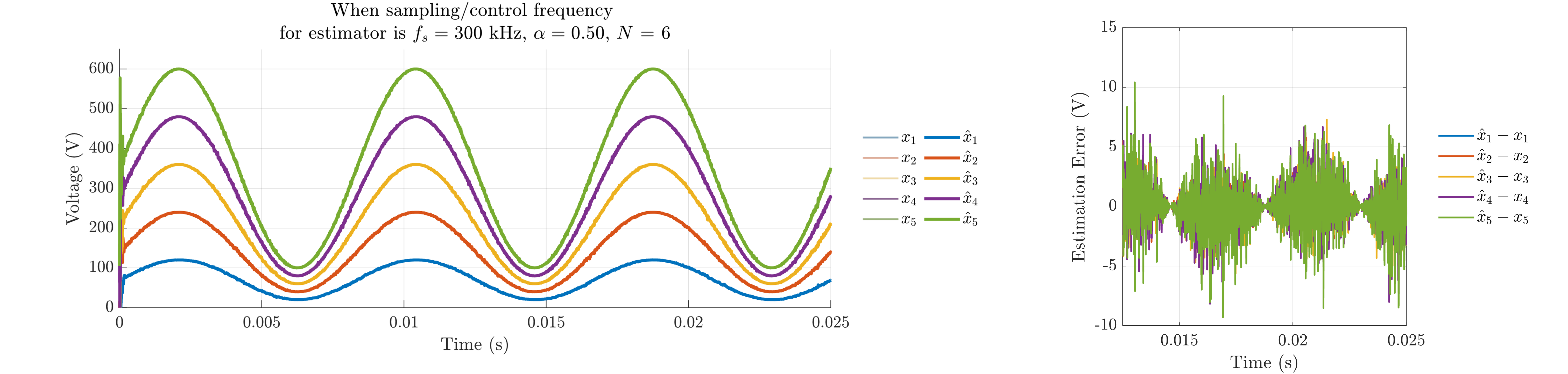} 
        \caption{$f_s = 300$ kHz, $\alpha = 0.5$, $N=6$}
        \label{fig:subfig1}
    \end{subfigure}
    \hfill 
    
    \begin{subfigure}[t]{0.7\linewidth} 
        \centering
        \includegraphics[width=\linewidth]{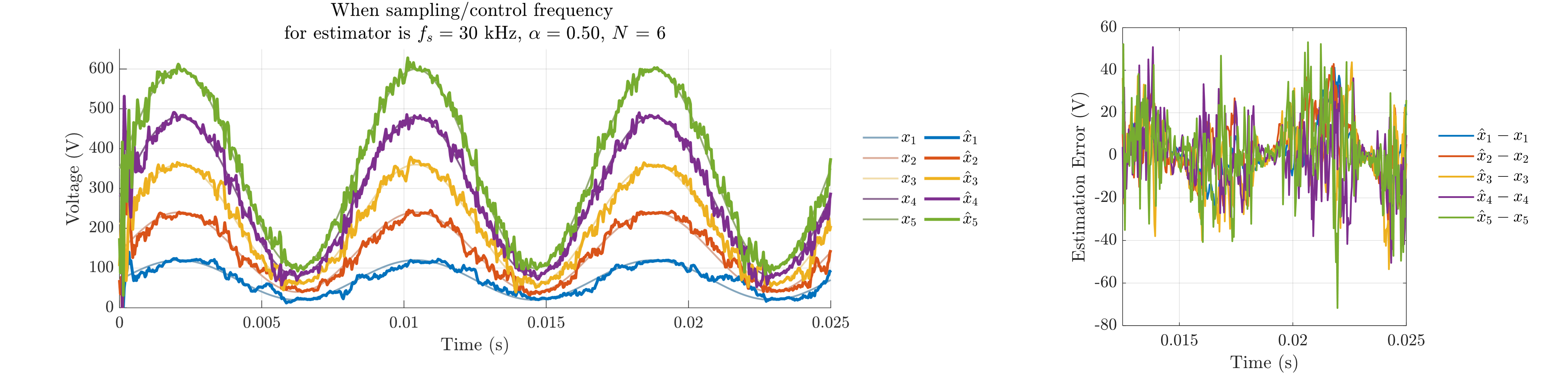} 
        \caption{$f_s = 30$ kHz, $\alpha = 0.5$, $N=6$}
        \label{fig:subfig2}
    \end{subfigure}
    \hfill
    
    \begin{subfigure}[t]{0.7\linewidth} 
        \centering
        \includegraphics[width=\linewidth]{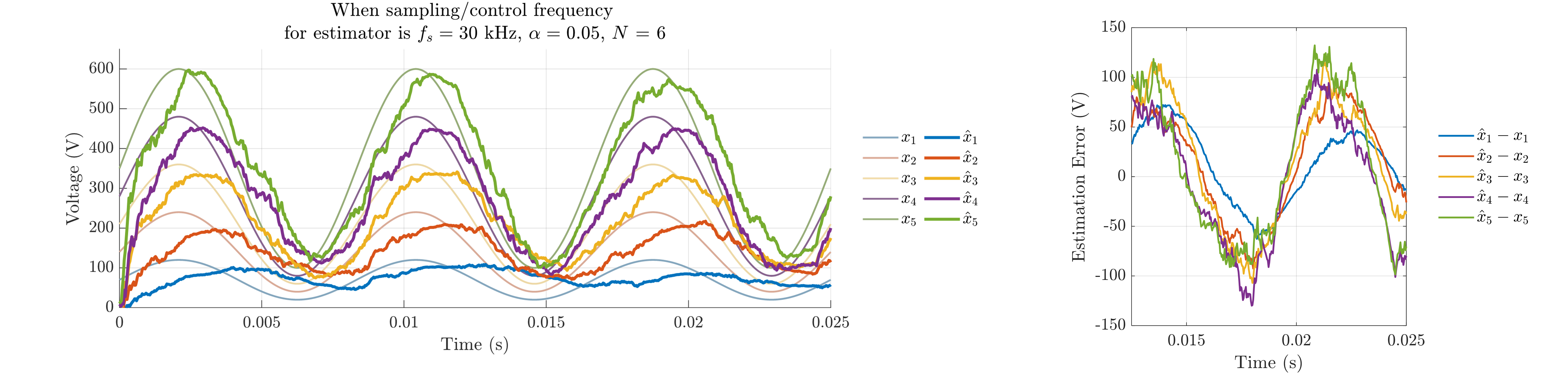} 
        \caption{$f_s = 30$ kHz, $\alpha = 0.05$, $N=6$}
        \label{fig:subfig3}
    \end{subfigure}
    \hfill
    
    \begin{subfigure}[t]{0.7\linewidth} 
        \centering
        \includegraphics[width=\linewidth]{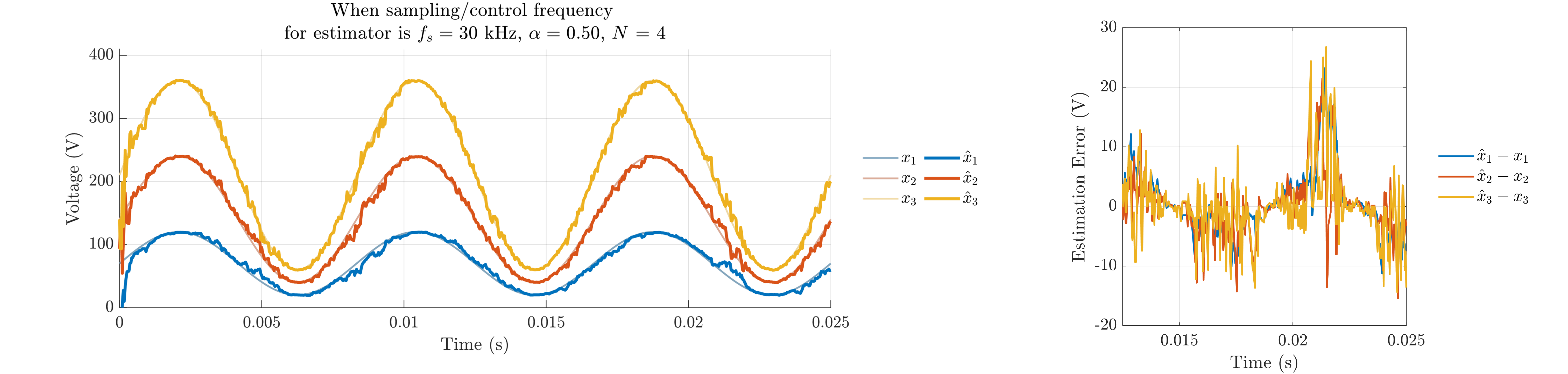} 
        \caption{$f_s = 30$ kHz, $\alpha = 0.5$, $N=4$}
        \label{fig:subfig4}
    \end{subfigure}
    
    \caption{Real-time estimation simulation result using a feedback-only estimator. As \( N \) increases, the bandwidth decreases sharply, leading to higher high-frequency errors and increased estimation errors at 120 Hz. A larger \( \alpha \) results in a higher average bandwidth, reducing the 120 Hz AC signal estimation error, but significantly increases instantaneous high-frequency errors at the sampling frequency level. Increasing the sampling frequency improves the bandwidth and significantly reduces overall errors.}
    \label{fig:mainfig}
\end{figure*}

According to (\ref{eig_range}), feedback gain ($\alpha$) should satisfy following inequality:
\begin{equation}
\alpha<\frac{2}{N-2}
\label{eig_range}
\end{equation} which at least ensures the stability of the estimator. Fig. \ref{alpha_fig} shows the impact of $\alpha$ on the stability of the feedback estimator and the mathematical properties of closed-loop estimation.
\subsection{Frequency Response Characteristics}
\eqref{system_matrix} can be expressed as follows:
\begin{equation}
\label{digital_domain}
\begin{aligned}
\frac{\mathbf{\hat{v_{c}}}\left[ n \right] - \mathbf{\hat{v_{c}}}\left[ n-1 \right]}{\tau_{s}}
= \, \mathbf{K_{est}}[n]  \left( \mathbf{v_{c}}\left[ n \right] - \mathbf{\hat{v}_{c}}\left[ n-1 \right] \right) &\\ + i_{L}[n] \frac{\mathbf{\Delta d^{*}}[n-1]}{\mathbf{C_{f}}}
\end{aligned}
\end{equation}
The feedback gain matrix in this formula is
\begin{equation}
\begin{aligned}
\mathbf{K_{est}}[n] = \frac{\alpha}{\tau_s} \mathbf{\Delta S}\left[ n \right] \Delta \mathbf{S}{{\left[ n \right]}^{\mathbf{T}}}
\end{aligned}
\label{var_gain}
\end{equation}
which varies at every sampling instants.
If the sampling frequency is significantly higher than the estimator's bandwidth, the discrete-time update function in \eqref{digital_domain} can be approximated in continuous-time domain. The equivalent representation in continuous-time domain is expressed as follows:
\begin{equation}
\label{cont}
\mathbf{\hat{v}_{c}}(t) \approx \int_{-\infty}^{t}{{\mathbf{K}}_{\mathbf{est}}}(\tau)\left( \mathbf{v_{c}}\left( \tau \right)-\mathbf{\hat{v}_{c}}\left( \tau \right) \right)+i_{L}(\tau)\frac{\mathbf{\Delta d^{*}(\tau)}}{\mathbf{C_f}} \, d\tau
\end{equation}

Here, the proposed estimator contains variable proportional gain term, $\mathbf{K_{est}}$, and feedforward term. The laplace transform of \eqref{cont} is as follows:
\begin{equation}
\label{s_domain}
\mathbf{\hat{V}_{c}}\left( s \right) \approx \frac{{{\mathbf{K}}_{\mathbf{est,eff}}}}{s}\left( \mathbf{V_{c}}\left( s \right)-\mathbf{\hat{V}_{c}}\left( s \right) \right)+{{{\mathbf{\hat{V}}}}_{\mathbf{ff}}}\left( s \right)
\end{equation}
where
\begin{equation}
\mathbf{K_{est,eff}}(s) = \frac{\mathbf{K_{est}}(s)*(\mathbf{V_c}(s)-\mathbf{\hat{V}_c}(s))}{\mathbf{V_c}(s)-\mathbf{\hat{V}_c}(s)}
\end{equation}
\begin{equation}
    \mathbf{\hat{V}_{ff}}(s)=\frac{1}{s}\cdot \mathcal{L}\{ {i_{L}(t)\frac{\mathbf{\Delta d^{*}(t)}}{\mathbf{C_f}}} \}(s)
\end{equation}
Then, \eqref{s_domain} can be expressed with following forms:
\begin{equation}
\label{fb_ff}
{{\mathbf{\hat{V}}}_{\mathbf{c}}}(s)=\underbrace{\frac{ {{\mathbf{K}}_{\mathbf{est,eff}}}(s)}{s}\left( {{{\mathbf{V}}}_{\mathbf{c}}}(s)-{{{\mathbf{\hat{V}}}}_{\mathbf{c}}(s)} \right)}_{feedback}+\underbrace{\mathbf{\hat{V}_{ff}}(s)}_{feedforward}
\end{equation}
\begin{equation}
\begin{split}
\label{lf_hf}
{{\mathbf{\hat{V}}}_{\mathbf{c}}(s)} &= \underbrace{\frac{{{\mathbf{K}}_{\mathbf{est,eff}}} }{ s\mathbf{I}+{{\mathbf{K}}_{\mathbf{est,eff}}} }{{\mathbf{V}}_{\mathbf{c}}(s)}}_{{low-pass-filter}} \\
&+ \underbrace{\frac{ s\mathbf{I} }{ s\mathbf{I}+{{\mathbf{K}}_{\mathbf{est,eff}}} }{\mathbf{\hat{V}_{ff}}(s)}}_{{high-pass-filter}}
\end{split}
\end{equation}

\begin{figure}[t]
    \centering
    \includegraphics[width=1\linewidth]{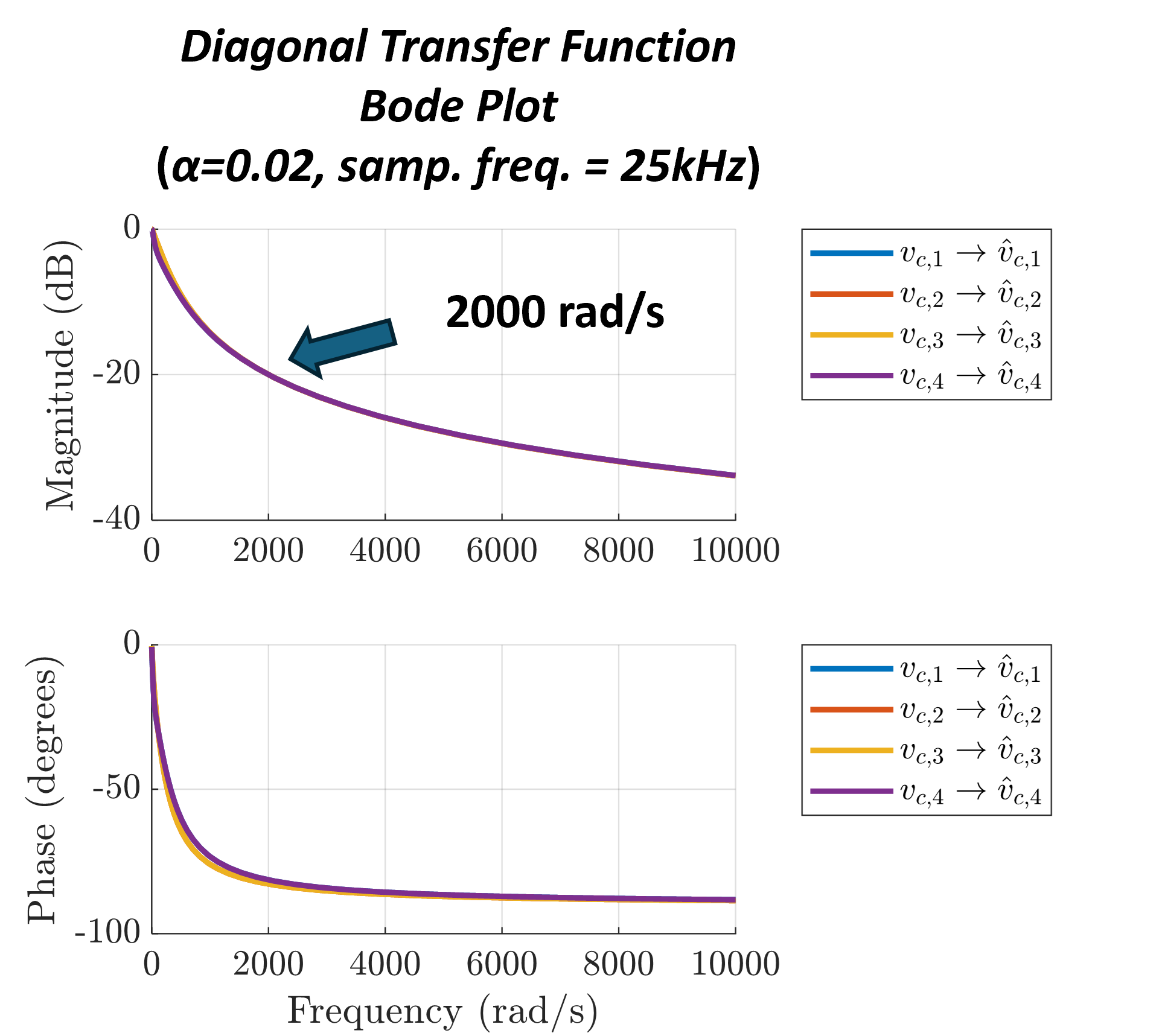}
    \caption{Bode plots of diagonal transfer functions for closed-loop estimator, where $f_s = 25$ kHz, $\alpha = 0.02$, and $\mathbf{d} = [0.42, 0.37, 0.41, 0.46, 0.5]^\mathbf{T}$. Even if the bandwidth of diagonal transfer function has high-bandwidth, non-diagonal term makes the effective bandwidth much lower.}
    \label{fig:Diag_tf}
\end{figure}

\begin{figure}[t]
    \hfill 
    \begin{subfigure}[b]{0.9\linewidth} 
        \centering
        \includegraphics[width=\linewidth]{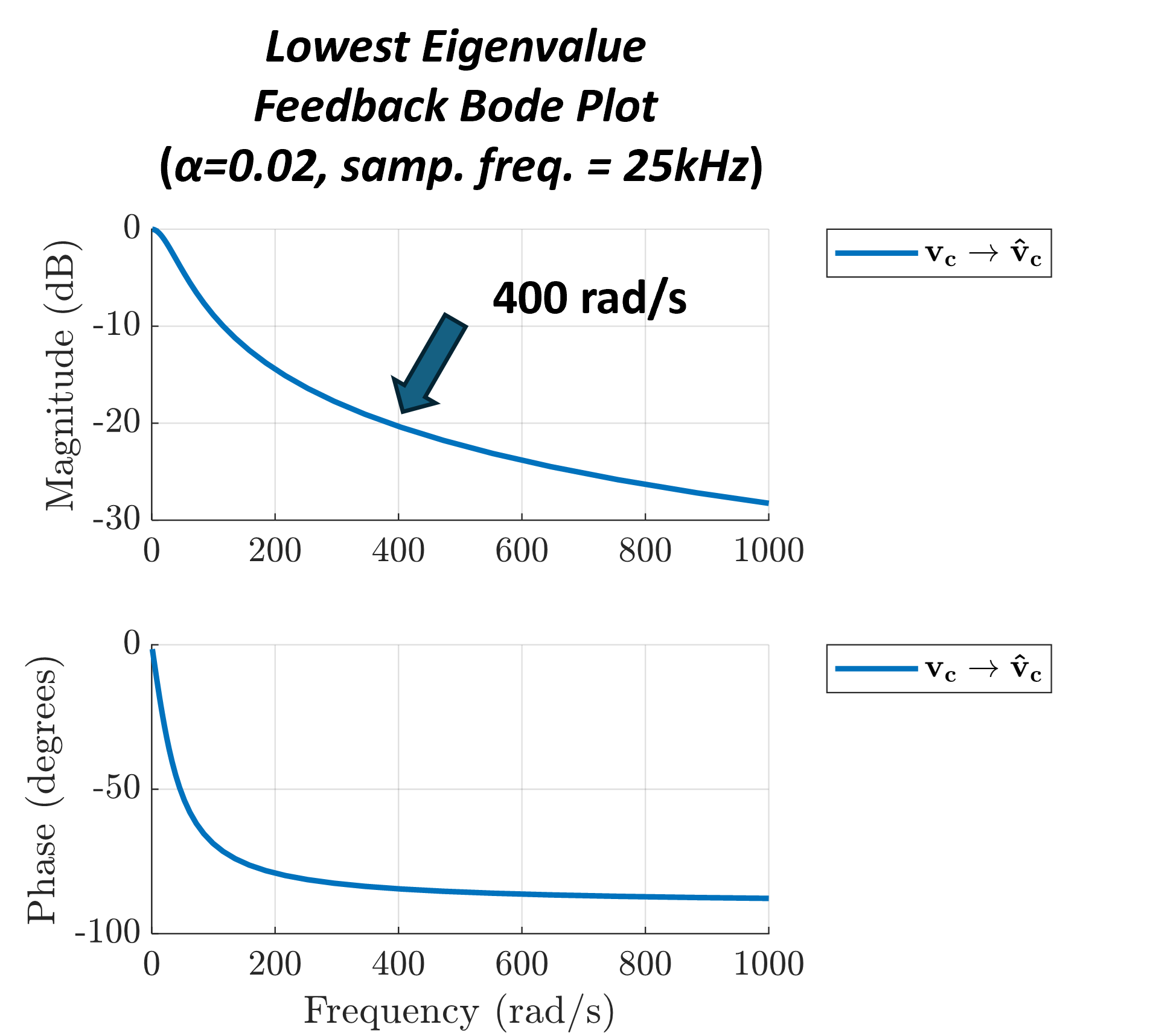} 
        \caption{$\alpha=0.02$}
    \end{subfigure}

    \hfill 
    \begin{subfigure}[b]{0.9\linewidth} 
        \centering
        \includegraphics[width=\linewidth]{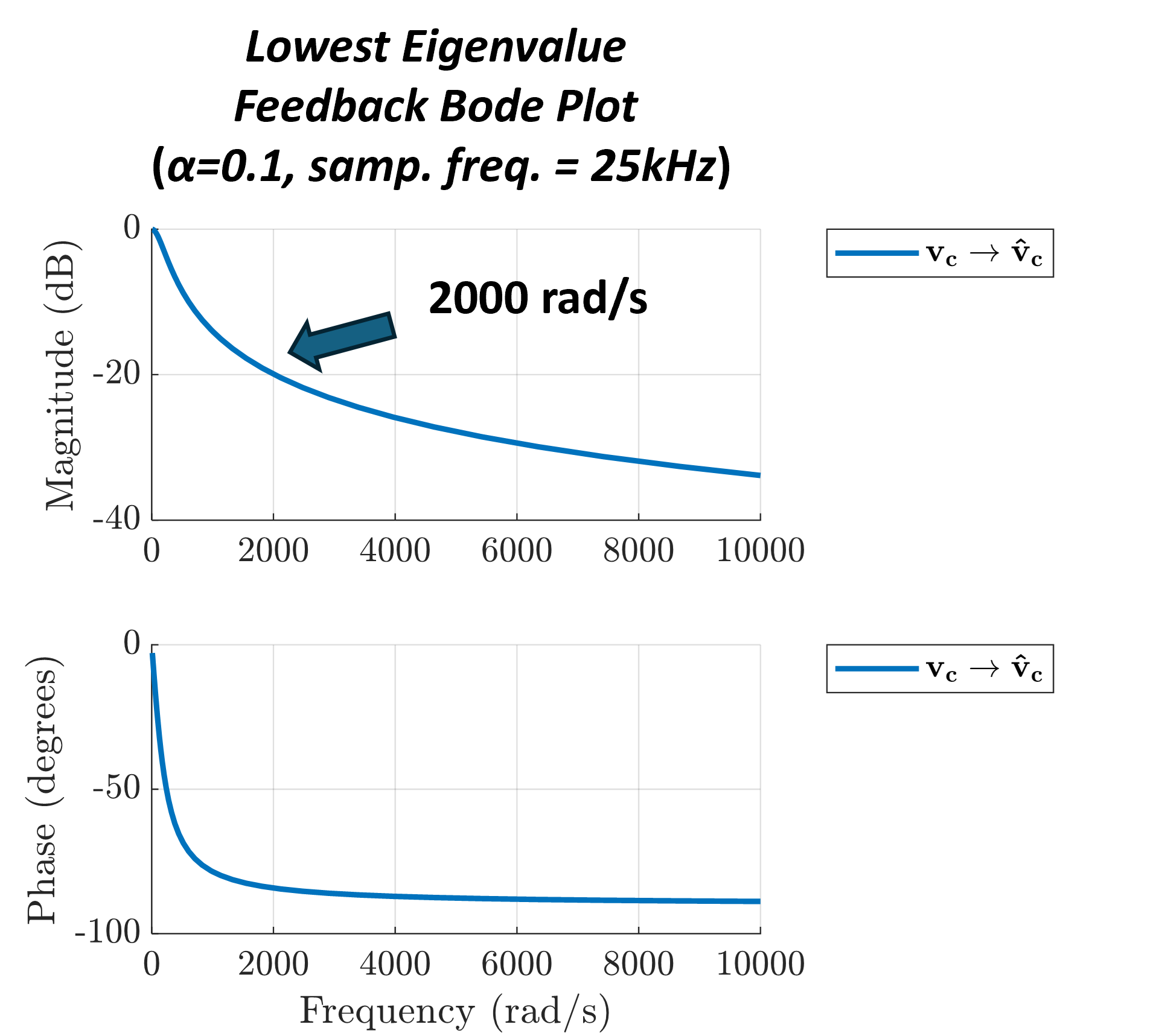} 
        \caption{$\alpha=0.1$}
    \end{subfigure}

    \caption{Bode plots of transfer functions for closed-loop estimator considering maximum eigenvalue of $\mathbf{P_{fr}}$, where $f_s = 25$ kHz and $\mathbf{d} = [0.42, 0.37, 0.41, 0.46, 0.5]^\mathbf{T}$. Compared to bode plot of diagonal transfer function in Fig.~\ref{fig:Diag_tf}, the bandwidth is much lower. As $\alpha$ increases, the bandwidth gets higher.}
    \label{fig:beta_bode}
\end{figure}
\begin{figure*}[t]
    \centering
    \includegraphics[width=1\linewidth]{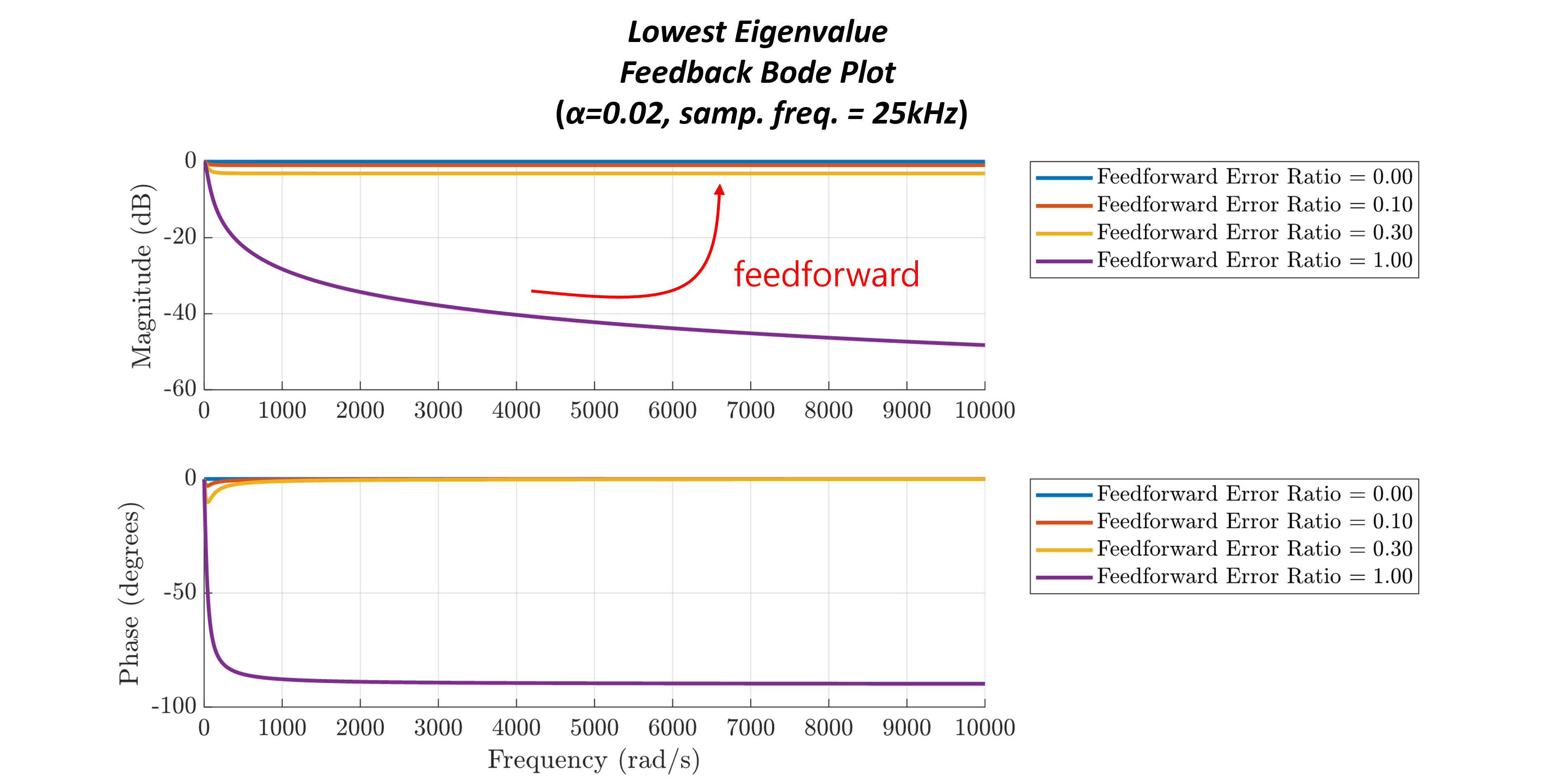}
    \caption{Bode plot of the hybrid estimator considering maximum eigenvalue of $\mathbf{P_{fr}}$, where $f_s = 25$ kHz and $\mathbf{d} = [0.42, 0.37, 0.41, 0.46, 0.5]^\mathbf{T}$. Compared to the feedback estimator shown in Fig.~\ref{fig:beta_bode}(a), the bandwidth and phase margin is highly improved. These improvements enhance a stability when used in cascaded connection with controllers. The 'Feedforward Error Ratio' represents deviations caused by various error components in the state feedforward term. A value of 1 indicates the feedback-only case.
}
    \label{fig:beta_tf_low_ff}
\end{figure*}
Fig.~\ref{fig:mainfig} shows the closed-loop estimation performance according to the FCML level ($N$), feedback gain ($\alpha$), and sampling/control frequency ($f_s$). In Fig.~\ref{fig:mainfig}(a), when \(f_s\) is high at 300 kHz, the estimation error is small. However, Fig.~\ref{fig:mainfig}(b) shows the estimation error increases significantly when \(f_s\) is reduced to 30 kHz. This error includes both high-frequency errors at the sampling frequency and errors in the fundamental frequency (120 Hz). As shown in Fig.~\ref{fig:mainfig}(c) the feedback gain (\(\alpha\)) is reduced to 0.05 compared to Fig.~\ref{fig:mainfig}(b), leading to a lower bandwidth and significantly degraded estimation performance for the 120 Hz signal. Here, the estimation delay increases as bandwidth lowered. Conversely, the high-frequency error at the sampling frequency is decreased in Fig.~\ref{fig:mainfig}(c) compared to Fig.~\ref{fig:mainfig}(b) due to lowered $\alpha$. Interestingly, when the FCML level (\(N\)) is reduced to 4, decreasing the number of variables to estimate to 3, the estimation error is significantly lowered in Fig.~\ref{fig:mainfig}(d) compared to the case in Fig.~\ref{fig:mainfig}(b), even with the same feedback gain and sampling frequency. This is because a lower \(N\) requires fewer \(N_{dis}\) to satisfy full-rank operation, reducing the impact of the single-rank of the multi-cost matrix on estimation error. 

In summary, the closed-loop estimation performance degrades significantly with lower sampling frequency and higher FCML level. While increasing \(\alpha\) improves the estimation bandwidth, it also increases the high-frequency error caused by the single-rank of the multi-cost matrix at the sampling frequency level.

As shown in Fig.~\ref{fig:Diag_tf}, the feedback term in the estimator acts as a low-pass filter as discussed in \eqref{lf_hf}. It is important to note that Fig.~\ref{fig:Diag_tf} utilizes the diagonal transfer function, while the actual frequency response has a much lower bandwidth shown in Fig.~\ref{fig:Diag_tf}. This is because, in a MIMO system, the bandwidth is determined by the lowest eigenvalue of the system matrix which cannot be shown in diagonal terms. A Bode plot of the transfer function, conservatively defined by the lowest eigenvalue, is shown in Fig.~\ref{fig:beta_bode}. It reveals that as \(\alpha\) increases, the bandwidth also increases, but remains significantly lower than that of the diagonal transfer function in Fig.~\ref{fig:Diag_tf}.

On the other hand, the feedforward term in \eqref{lf_hf}, derived from the open-loop estimation, behaves as a high-pass filter, estimating high-frequency components and rapid variations. 

If the parameter error is zero and sampled inductor current and applied duty is same with actual values in ideal case, then, following equality holds:
\begin{equation}
{{\mathbf{V}}_{\mathbf{c}}}(s)\approx \frac{1}{s} \cdot \mathcal{L}\{\frac{{{i}_{L}}\mathbf{\Delta }{{\mathbf{d}}^{*}}}{{{{\mathbf{C}}}_{\mathbf{f}}}}\}(s)
\end{equation}
therefore, 
\begin{equation}
\frac{{{{\mathbf{\hat{v}}}}_{\mathbf{c}}}}{{{\mathbf{v}}_{\mathbf{c}}}}\approx \mathbf{1}
\end{equation} which shows that the estimated value is exactly same with the actual value, without any estimation delay and estimation errors. The Bode plot of the hybrid estimator with feedforward is shown in Fig.~\ref{fig:beta_tf_low_ff}. This shows the performance of the hybrid estimator with fast dynamics. However, in real world, the parameter/sampling error can occur, which generates the feedforward term has some error. Despite minor errors in feedforward term, Fig.~\ref{fig:beta_tf_low_ff} shows the the feedforward term enhances the estimator's performance.

In summary, the proposed hybrid estimator separates estimation tasks: low-frequency components are handled by closed-loop feedback, while high-frequency dynamics are managed by an open-loop feedforward path. The feedforward's high-pass filtering characteristic prevents integrator wind-up by rejecting DC errors and allowing only high-frequency dynamics. This design achieves high bandwidth without requiring excessively high sampling rates, simplifying implementation while maintaining strong dynamic performance. In contrast, conventional feedback-only methods act as low-pass filters and require a high feedback gain (\(\alpha\)) to minimize delay. The impact of feedforward errors, dependent on the choice of \(\alpha\), will be discussed in the next chapter.

\subsection{Gain Setting}
\label{bounds}
The proposed estimator employs an effective feedback gain matrix, $\mathbf{K_{\text{est}}}[n]$, as defined in \eqref{var_gain}, which varies with the switching states. Furthermore, the feedback matrix depends on the FCML level and the combinations of duty cycle references. Due to this variability, $\alpha$ must be configured to account for the worst-case scenarios, averaged over $N_{\text{dis}}$ sampling instants, as illustrated in Fig.~\ref{fig:beta_bode}. If the FCML level and operating conditions are constrained, the gain can be adjusted more flexibly, enabling improved performance under specific conditions.

\subsubsection{Upper Bound, High Frequency Error from Instantaneous Rank Deficiency}
\indent The proposed estimator utilizes a multi-cost gradient descent approach because the exact gradient of the cost function in (\ref{opt}) cannot be directly obtained under the given conditions. However, it entails estimation errors oscillating at the sampling frequency as shown in Fig.~\ref{fig:mainfig}. These errors arise from discrepancies between the actual gradient of the cost function and the switching state vector, caused by a single rank of system matrix at each time instant.

The system matrix equation in \eqref{system_matrix} can be rewritten as:
\begin{equation}
\begin{aligned}
\mathbf{\hat{v}_{c}}[n] - \mathbf{\hat{v}_{c}}[n-1] &= \mathbf{\Delta \hat{v}_{fb}}[n] + \mathbf{\Delta \hat{v}_{ff}}[n],
\end{aligned}
\end{equation}
where
\begin{equation}
\mathbf{\Delta \hat{v}_{fb}}[n] = \alpha \mathbf{\Delta S}[n] \mathbf{\Delta S}[n]^\mathbf{T} (\mathbf{v_{c}}[n] - \mathbf{\hat{v}_{c}}[n-1]) 
\end{equation}

To mitigate the high-frequency estimation error from the feedback term, the feedback gain ($\alpha$) must be carefully selected. A high $\alpha$ accelerates the feedback response but amplifies high frequency errors, potentially introducing disturbances in both current control and active voltage balancing.

\indent The variation in the estimated value ($||\mathbf{\Delta \hat{v}_{fb}}[n]||_{\infty}$) updated by the feedback term must be controlled to reduce excessive high-frequency noise. To achieve this, assuming that the estimation error at $t = (n-1)\tau_s$ is zero, the following inequality can be applied:
\begin{equation}
\begin{aligned}
||\mathbf{\Delta \hat{v}_{fb}}[n]||_{\infty} &= \alpha ||\mathbf{\Delta S}[n] \mathbf{\Delta S}[n]^\mathbf{T} (\mathbf{v_{c}}[n] - \mathbf{\hat{v}_{c}}[n-1])||_{\infty} \\
&\leq \alpha ||\mathbf{\Delta S}[n]||_{\infty}|| \mathbf{\Delta S}[n]^\mathbf{T}(\mathbf{v_{c}}[n] - \mathbf{\hat{v}_{c}}[n-1])||_{\infty} \\
&\leq \alpha  \mathbf{1^{T}}
 |\mathbf{v_{c}}[n] - \mathbf{\hat{v}_{c}}[n-1]| \\
& \leq \alpha\tau_s \sum_{k=1}^{N-2}\frac{k}{N-1}\max(\frac{dv_{in}}{dt}) \\
& =\alpha\tau_s \frac{N-2}{2}\max(\frac{dv_{in}}{dt})
\leq v_{e,HF}
\end{aligned}
\end{equation} where $v_{e,HF}$ is the allowable maximum high-frequency error from feedback term. The upper bound of $\alpha$ can be set as follows:
\begin{equation}
\alpha
\leq \frac{2v_{e,HF}}{\tau_s (N-2) \max(\frac{dv_{in}}{dt})}
\label{alpha_u_bound}
\end{equation}

\indent As the sampling period ($\tau_s$) decreases, indicating higher sampling and control frequencies, the upper bound of $\alpha$ in (\ref{alpha_u_bound}) increases. This tendancy can be found in Fig.~\ref{fig:mainfig}. This is because the voltage difference of the input voltage $(\frac{dv_{in}}{dt})$ is reduced with faster sampling. Conversely, as the level of the FCML increases, the upper bound decreases. This occurs because the dimension of the vector space for flying capacitor voltages grows with higher levels, requiring more sampling instances ($N_{dis}$) to achieve full-rank operation. 

In summary, the instantaneous rank deficiency of the multi-cost matrix introduces high frequency errors during a single update. These errors become more significant as the FCML level increases, given that achieving full-rank operation requires $(N-2)$ dimensions.

\subsubsection{Lower Bound, Step 1: Eigenvalues of the System Matrix}

\indent The matrix 
\begin{equation}
\label{R}
\mathbf{R} = 
\left[ \mathbf{\Delta S}[n_0+1], \mathbf{\Delta S}[n_0+2], \dots, \mathbf{\Delta S}[n_0+N_{dis}] \right] 
\end{equation}
contains the eigenvectors of each system matrix \(\mathbf{P}[n]\). For ensuring observability, the matrix $\mathbf{R}$ must achieve full rank (\(N-2\)).

Meanwhile, the eigenvalues of the product matrix 
\(\mathbf{P_{fr}} = \mathbf{P}[n_0+N_{dis}] \mathbf{P}[n_0+N_{dis}-1] \dots \mathbf{P}[n_0+1]\) 
are given as:
\begin{equation}
\label{prod_P}
\operatorname{eigval}(\mathbf{P_{fr}}) =  \operatorname{eigval}(\prod_{i=N_{dis}}^{1}\mathbf{P}[n_0+i]).
\end{equation} The eigenvectors/eigenspace and eigenvalues of each individual system matrix (\(\mathbf{P}[n]\)) are defined as:
\begin{equation}
\label{eigv}
\{\mathbf{\Delta S}[n], \mathbb{R}^{N-2} \setminus \text{span}\{\mathbf{\Delta S}[n]\}\} \in \operatorname{eigvec}(\mathbf{P}[n]),
\end{equation}
\begin{equation}
\label{eig}
\{1-\alpha \mathbf{\Delta S}[n]^\mathbf{T}\mathbf{\Delta S}[n], 1\} \in \operatorname{eigval}(\mathbf{P}[n]),
\end{equation}, respectively. Here, the eigenvalue \(1\) has \(N-3\) degeneration. All eigenvalues of \(\mathbf{P}[n]\) lie within the range \((-1, 1]\) according to \eqref{prod_P}, \eqref{eig}.\\
\indent However, for the product matrix \(\mathbf{P_{fr}}\), all eigenvalues lie strictly within the range \((-1, 1)\). If an eigenvalue of \(\mathbf{P_{fr}}\) equaled \(1\), the following condition must have held:
\begin{equation}
\mathbf{P_{fr}} \mathbf{x} = \mathbf{x} \, \, \text{for} \, \, \exists \mathbf{x} \in \mathbb{R}^{N-2}.
\end{equation}
This condition implies that the eigenvector \(\mathbf{x}\), corresponding to the eigenvalue \(1\), resides in the orthogonal complement of the span of 
\(\{\mathbf{\Delta S}[n_0+1], \mathbf{\Delta S}[n_0+2], \dots, \mathbf{\Delta S}[n_0+N_{dis}]\}\), according to \eqref{R} and \eqref{eigv}. However, since \(\mathbf{P_{fr}}\) has full rank, the span fully covers \(\mathbb{R}^{N-2}\). Consequently, no eigenvector \(\mathbf{x}\) exists that satisfies this condition. Therefore, all eigenvalues of \(\mathbf{P_{fr}}\) are strictly confined to the range \((-1, 1)\).

\indent The maximum eigenvalue of \(\mathbf{P_{fr}}\) can be expressed as:
\begin{equation}
\max(\operatorname{eigval}(\mathbf{P_{fr}})) = \beta_{\text{max}} < 1.
\end{equation}
As \(\alpha\) increases, \(\beta_{\text{max}}\) decreases strictly as shown in Fig.~\ref{beta_lower}. This behavior shows the influence of \(\alpha\) on the eigenvalue spectrum of \(\mathbf{P_{fr}}\). A higher value of \(\alpha\) results in a lower \(\beta_{\text{max}}\), leading to faster convergence of the estimator.

\subsubsection{Lower Bound, Step 2. The Effect of Parameter and Sampling Error}
\begin{figure}[t]
    \centering
    \includegraphics[trim=2cm 9cm 2cm 9cm, clip, width=1\linewidth]{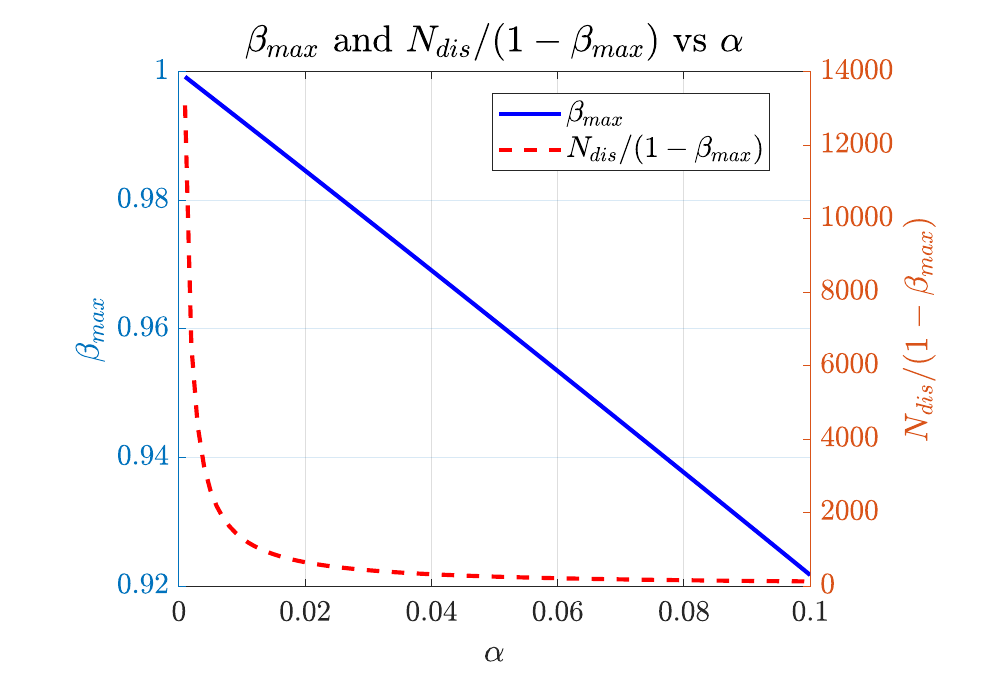}
    \caption{Graph showing the variation of \( \beta_\text{max} \) and \( N_\text{dis}/(1-\beta_\text{max}) \) as functions of \( \alpha \), where \( N=6 \), \( f_s = 25 \, \text{kHz} \), and \( \mathbf{d} = [0.42, 0.37, 0.41, 0.46, 0.5]^\mathbf{T} \). As \( \alpha \) increases, the \( \alpha \)-dependent \( \beta_\text{max} \) decreases, leading to a reduction in \( N_\text{dis}/(1-\beta_\text{max}) \), which affects the DC offset error. Consequently, a small \( \alpha \) results in a significantly larger DC offset error.}
    \label{beta_lower}
\end{figure}

For simpicity, the extra term except for the estimation error component, $\mathbf{\tilde{v}}_{c} = \mathbf{\hat{v}}_{c} - \mathbf{{v}}_{c}$, can be expressed with $\mathbf{u}[n]$ as follows:
\begin{equation}
\begin{aligned}
&
  \mathbf{u}\left[ n \right]=\alpha \mathbf{\Delta S}\left[ n \right]\mathbf{\Delta S}{{\left[ n \right]}^{\mathbf{T}}}\\&\times\left( {{v}_{c}}\left[ n \right]-{{v}_{c}}\left[ n-1 \right] \right)+\tau_{s}\mathbf{\Delta }{{{\mathbf{\tilde{v}}}}_{\mathbf{ff}}}
  \end{aligned}
\end{equation}Then \eqref{system_matrix} can be simplified as follows:
\begin{equation}
  {{{\mathbf{\tilde{v}}}}_{\mathbf{c}}}\left[ n \right]=\mathbf{P}\left[ n \right]{{{\mathbf{\tilde{v}}}}_{\mathbf{c}}}\left[ n-1 \right]+\mathbf{u}\left[ n \right]
\end{equation}
With this, the estimation error is calculated as follows:
\begin{equation}
  {{{\mathbf{\tilde{v}}}}_{\mathbf{c}}}\left[ n \right]=\left( \prod\limits_{l=1}^{n}{\mathbf{P}\left[ l \right]} \right)\mathbf{\tilde{v}_{c}}\left[ 0 \right]+\sum\limits_{l=0}^{n}{\left( \prod\limits_{k=l+1}^{n}{\mathbf{P}\left[ k \right]} \right)\mathbf{u}\left[ l \right]}
\end{equation}
The upper bound of DC-error in steady state condition is calculated as follows:
\begin{equation}
\begin{aligned}
  \lim_{n \to \infty} \left\| {{{\mathbf{\tilde{v}}}}_{\mathbf{c}}}\left[ n \right] \right\|_{\infty} 
  &= \lim_{n \to \infty} \left\| \left( \prod_{l=1}^{n} \mathbf{P}\left[ l \right] \right) \mathbf{\tilde{v}_{c}}\left[ 0 \right] \right. \\
  & \quad + \left. \sum_{l=0}^{n} \left( \prod_{k=l+1}^{n} \mathbf{P}\left[ k \right] \right) \mathbf{u}[l] \right\|_{\infty} \\
  &= \underset{n\to \infty }{\mathop{\lim }}\,{{\left\| \sum\limits_{l=0}^{n}{\left( \prod\limits_{k=l+1}^{n}{\mathbf{P}\left[ k \right]} \right)\mathbf{u}\left[ l \right]} \right\|}_{\infty }}\\
  &\le {{N}_{dis}}\frac{{{\left\| \mathbf{u}[l] \right\|}_{\infty }}}{1-{{\beta }_{\max }}}\\
  &\approx {{N}_{dis}}\tau_{s}\frac{{{\left\| \mathbf{\Delta {{{\tilde{v}}}_{ff}}} \right\|}_{\infty }}}{1-{{\beta }_{\max }}}
\end{aligned}
\end{equation}
As discussed before, $\beta_{\text{max}}$ is strictly decreased as $\alpha$ increased, meaning that as $\alpha$ increases, the DC-offset error in the estimation value from the feedforward error (${{{\left\| \mathbf{\Delta {{{\tilde{v}}}_{ff}}} \right\|}_{\infty }}}$) increases, which implies that $\alpha$ must be sufficiently large to minimize the DC-offset error. \noindent Accordingly, the lower bound of $\alpha$ is derived as follows:
\begin{equation}
\alpha \geq \beta^{-1}_{\text{max}}\left(1 - N_{dis}\tau_s \frac{{{\left\| \mathbf{\Delta \tilde{v}_{ff}} \right\|}_{\infty}}}{v_{e,DC}}\right)
\end{equation}
where $\beta^{-1}_{\text{max}}$ is a decreasing function as shown in Fig.~\ref{beta_lower}, and ${v_{e,DC}}$ is allowable maximum DC estimation error. As the FCML level ($N$) increases, $N_{\text{dis}}$ also rises according to \eqref{N_dis}, leading to a higher lower bound for $\alpha$ to maintain the same DC offset error. This is caused by the instantaneous rank-deficiency of the system matrix ($\mathbf{P}[n]$).

A larger feedback gain ($\alpha$) reduces the DC offset error caused by feedforward inaccuracies. However, as $N$ increases, the rank-deficiency effect becomes more severe, further raising the minimum required $\alpha$. Conversely, higher sampling frequencies (shorter sampling periods) alleviate this problem by resolving rank-deficiency more quickly.

\subsubsection{Summary of Gain Setting}
\noindent Both the upper and lower bounds for $\alpha$ are influenced by the FCML level ($N$) and the sampling frequency. Higher $N$ and lower sampling frequencies increase the difficulty of selecting an appropriate gain due to amplified high-frequency and DC errors, which stem from the rank-deficiency effects of the multi-cost gradient descent method.

\begin{figure}[t]
    \centering
    \includegraphics[width=1\linewidth]{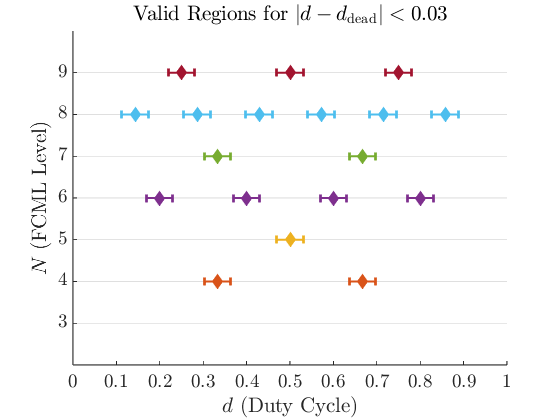}
    \caption{Graph showing the regions where sampled pole voltage information cannot be used for feedback due to switching effects. Around \( d_{dead} \), feedback is set to be disabled within a duty cycle margin of 0.03.}
    \label{valid}
\end{figure}

\begin{figure}[t]
    \centering
    \includegraphics[width=1\linewidth]{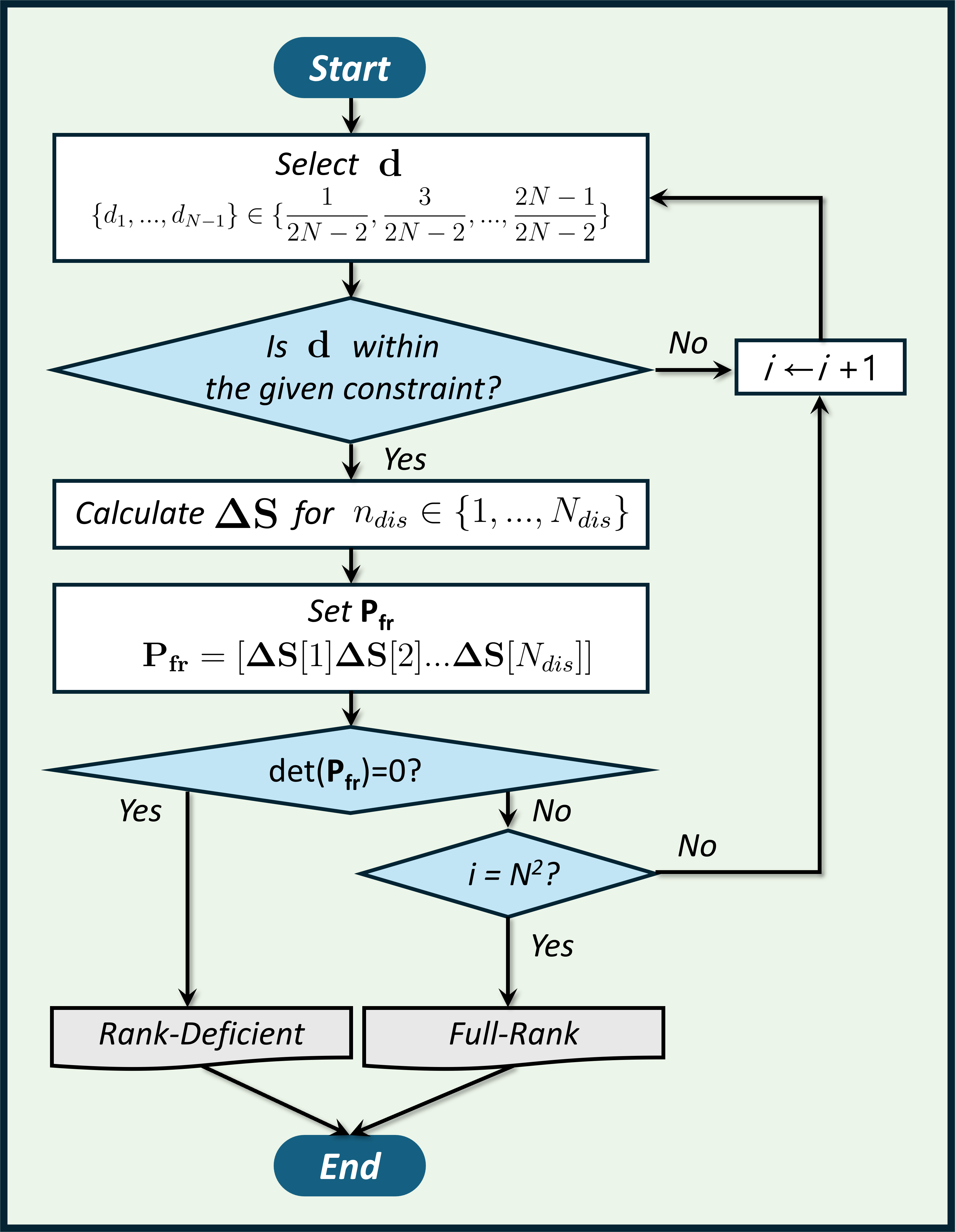}
    \caption{Algorithm to verify the feasibility of full-rank operation. The algorithm checks whether the disjoint sampling \( \mathbf{\Delta S} \) satisfies the full-rank condition across all duty cycle regions. When constraints are imposed on \( \mathbf{\Delta d} \) due to active balancing, the algorithm evaluates full-rank operation only within the valid duty cycle range.}
    \label{algo}
\end{figure}
\begin{figure*}[t]
    \centering
    \includegraphics[width=0.7\textwidth]{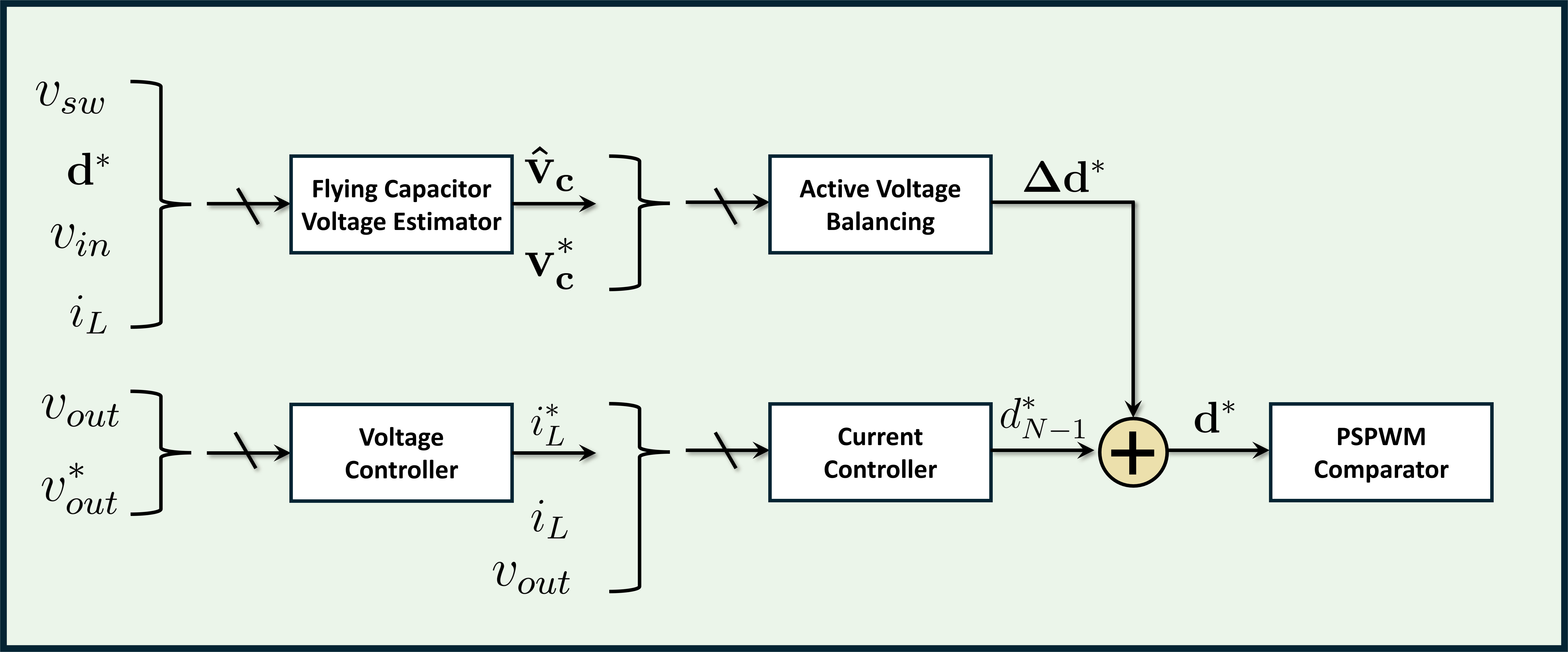}
    \caption{Block diagram of the utilized control system for estimator-based control.
}
    \label{fig:Controllers}
\end{figure*}
\begin{figure*}[t]
    \centering
    \includegraphics[width=0.8\linewidth]{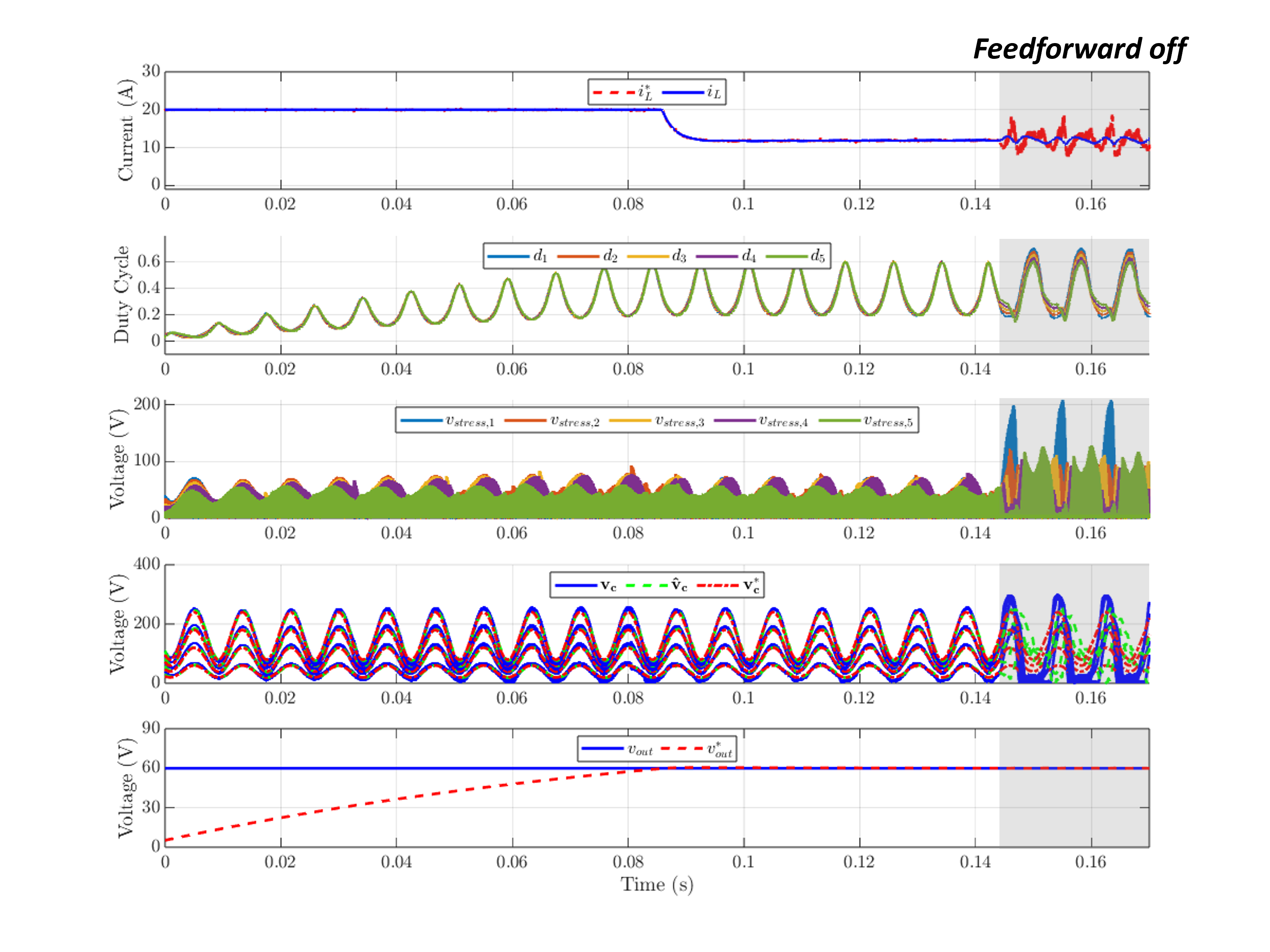}
    \caption{Results of estimator-based output capacitor voltage / active balancing / current control using the proposed hybrid estimator.}
    \label{fig:result}
\end{figure*}
\subsection{Prolonged Rank-Deficiency Problem}
\label{rank_def}
In certain scenarios, prolonged rank-deficiency problem can happen. This situation may cause the estimated value to diverge due to feedforward errors or oscillate without any updates in null-space of $\mathbf{P_{fr}}$ from the feedback term.

\subsubsection{Switching Noise at Sampling Instants}
During switching transitions, factors such as parasitic inductance in current commutation path, the rising/falling time caused by the gate-source capacitor's charging/discharging through the gate driver, and PWM signal delays can prevent the pole voltage from settling at the sampling instants. Consequently, the  unsettled pole voltage may be sampled instead. This results in distorted estimation through feedback term. This issue becomes particularly significant in DC-DC conversion cases where the duty cycle (\(\approx v_{out}/v_{in}\)) does not inherently change in steady state.

The duty cycle at which switching and sampling can coincide (\(d_{dead}\)) is calculated as follows:
\begin{equation}
    d_{dead} = 
    \begin{cases} 
      \dfrac{k}{N-1}, & \text{if } N \text{ is even}, \\[10pt]
      & k \in \{1, 2, \dots, N-2\}, \\[10pt]
      \dfrac{2k}{N-1}, & \text{if } N \text{ is odd}, \\[10pt]
      & k \in \left\{ 1, 2, \dots, \dfrac{N-3}{2} \right\}.
    \end{cases}
\end{equation}
For $N = 3 ~ N = 9$, $d_{dead}$ is depicted in Fig.~\ref{valid}. The duty cycles near \(d_{dead}\) can be also influenced by unsettled pole voltage. As a result, the proposed method has limitations in accurately estimating voltages due to unsettled pole voltage, as switching effects distort the sampling and feedback estimation. In such case, two methods can be considered to avoid the problem. 

First, dithering in the duty cycle reference can be used to avoid the unsettled pole voltage at sampling instants \cite{9829954}. However, this dithering introduces additional ripples in the inductor current and flying capacitor voltage due to changes in the effective pole voltage, which can also affect to their controllers. To address this, the controller can utilize anti-windup to eliminate the additionally induced current and voltage by the dithering.

\begin{table}[t]
\centering
\scalebox{1.3}{  
\begin{tabular}{|c|c|c|}
\hline
\textbf{N (Level of FCML)} & \textbf{Full-Rank?} & \textbf{$\Delta d_{\text{max}}$} \\ 
\hline
3 & $\bigcirc$ & 1 \\ 
\hline
4 & $\bigcirc$ & 1 \\ 
\hline
5 & $\times$ & 0 \\ 
\hline
6 & $\triangle$ & 0.2 \\ 
\hline
$N \geq 7$ & $\times$ & 0 \\ 
\hline
\end{tabular}
}
\caption{Table showing the feasibility of full-rank operation for various FCML levels, determined based on the algorithm described in Fig.~\ref{algo}. The symbol \( \triangle \) indicates cases where full-rank operation is achievable only under given constraints ($|\mathbf{\Delta d}|<\Delta d_{max}$).
}
\label{tbl_rank}
\end{table}
Secondly, by modifying the PWM method, the effective pole voltage can be maintained while preventing the switching instant and sampling instant from occurring simultaneously. By employing the skipped adjacency PWM (SAPWM) in \cite{hwang2024skippedadjacencypulsewidth}, the duty references can avoid the problematic region. However, this method requires additional external digital circuit and has the drawback of doubling the volt-second, leading to increase in switching ripple on inductor current and flying capacitor voltage.

In AC-DC buck operation, the periodic variation of the input voltage in steady state causes continuous changes in the required duty cycle, resulting in temporary rank-deficiency during specific time intervals. When rank-deficiency occurs due to switching effects, $\alpha$ can be temporarily set to zero, relying entirely on the feedforward term for estimation. This approach is more effective than using dithering or SAPWM, which introduce additional switching ripple. By temporarily utilizing the feedforward term, the rank-deficiency issue caused by switching effects can be effectively addressed, particularly in AC-DC buck operation.

\subsubsection{Insufficiency of Disjoint Sampling For Full-Rank Operation}
The proposed method relies on disjoint sampling for the pole voltage, requiring the switching state vectors to achieve full-rank operation across \(N_{dis}\) sampling instants. However, where active balancing is essential, the duty cycle references applied to each switch can differ, and the degree of freedom in duty combination increases with the FCML level ($N$). Consequently, it is necessary to evaluate whether the disjoint sampling is applicable for full-rank operation across the duty cycle combinations.\\
\indent Using the algorithm diagram provided in Fig.~\ref{algo}, the full-rank operation was iteratively verified in MATLAB for all duty cycle vectors. The result in TABLE~\ref{tbl_rank} reveals that for \(N \geq 5\), the disjoint sampling does not guarantee full-rank operation. Specifically, during the \(N_{dis}\) samplings, where peak/valley sampling of the PSPWM carrier is used, the set of the switching state vectors fail to satisfy full-rank operation for all duty cycle combinations. Therefore, the proposed method can only be applied universally to FCMLs with \(N = 3\) or \(N = 4\).\\
\indent However, in practical scenarios, $|\mathbf{\Delta d}|$ is typically limited as discussed in \eqref{lim_bal} to small values (e.g. 0.05) to limit the impact of active balancing controller on the current controller. These constraints restrict the duty cycle references generated by the active voltage balancing controller. When $|\mathbf{\Delta d}|$ is limited to 0.2 or less, full-rank operation becomes feasible even for \(N = 6\). As a note, in the case of \( N = 5 \), it is difficult to achieve a full rank operation through disjoint sampling compared to \( N = 6 \) because the peak and valley points of the PSPWM carriers overlap as shown in Fig.~\ref{fig:odd_even}.\\
\indent In summary, the proposed method is applicable to FCMLs with \(N = 3\), \(N = 4\), and \(N = 6\). 6-level FCML is particularly relevant in cases where estimator-based control is required for active balancing, especially in AC-DC buck operations. Notably, the constrained output of active balancing controller $(\mathbf{\Delta d})$ enables full-rank operation for \(N = 6\), making it suitable for grid-connected AC-DC buck converters employing 100 V GaN devices. This highlights the potential for data center applications, allowing low-cost CPUs to implement estimator-based control for 6-level FCML AC-DC buck conversion.\\
\indent Furthermore, to extend the applicability of the proposed method, additional voltage sensors can be employed to relax the full-rank condition. This enables full-rank operation for FCML with other levels \(N = 4\) or \(N \geq 7\). The placement of these voltage sensors should be optimized to effectively ensure full-rank operation, providing an efficient solution in terms of hardware design.
\section{Results}
Fig.~\ref{fig:result} presents the simulation results of the estimator-based control system, including output voltage control, active voltage balancing, and current control. The cascaded control system, as shown in Fig.~\ref{fig:Controllers}, is designed with time-scale separation to ensure non-interference among controllers. The FCML parameters and controller bandwidths are listed in TABLE~\ref{tab:fcml_parameters}. The simulation verifies the high-bandwidth characteristics of the proposed estimator by demonstrating DC current control for output voltage regulation under a varying input voltage of 120 Hz. The sampling frequency is approximately 25 kHz, suitable for low-cost MCUs.

During \(t = 0 \sim 0.145\) s, the output voltage is controlled from 0 V to 60 V, with the current reference limited to 20 A. The estimated flying capacitor voltage closely tracks the actual value, ensuring effective active voltage balancing. This prevents overvoltage, maintaining voltage stress on all switching devices below 100 V. As the output voltage increases, the required duty cycle also rises. Full-rank operation is maintained under the duty difference constraint, ensuring accurate estimation throughout the simulation. After the output voltage reaches the reference value, the current reference is reduced to match the load current without integrator wind-up issue, maintaining the output voltage.\\
\indent After \(t = 0.145\) s, the feedforward input is disabled to see the importance of state feedforward, leaving the estimator to rely solely on feedback. Due to the combination of low \(\alpha\), a high FCML level (\(N = 6\)), and a low sampling frequency, the feedback estimation significantly degrades, increasing estimation errors. This leads to poor current control and active voltage balancing, resulting in excessive voltage stress on switching devices, reaching nearly 200 V at maximum. Such overvoltage may lead to failure in 100 V-rated switching devices. The results highlight the importance of feedforward input and proper bandwidth settings for maintaining stable operation of all controllers. The result of the proposed method shows its superiority by enabling high-bandwidth estimator-based control even at low sampling rates. This highlights the effectiveness and practicality of the estimator in achieving robust control performance with reduced computational and sampling requirements.
 
\begin{table}[h!]
    \centering
    \begin{tabular}{|l|l|}
        \hline
        \textbf{Parameter} & \textbf{Value} \\ \hline
        FCML Level (\(N\)) & 6 \\ \hline
        Switching frequency (\(f_{\text{sw}}\)) & 120 kHz \\ \hline
        Effective switching frequency (\(f_{\text{sw,eff}}\)) & 600 kHz \\ \hline
        Sampling frequency (\(f_s\)) & 25.53 kHz (\(m_s = 47\)) \\ \hline
        Output voltage reference (\(v^{*}_{\text{out}}\)) & 60 V \\ \hline
        Input voltage frequency (\(2\omega_r\)) & 120 Hz \\ \hline
        Inductor (\(L\)) & 100 \(\mu\)H \\ \hline
        Output capacitor (\(C_{\text{out}}\)) & 20 mF \\ \hline
        Flying capacitor (\(C_d\)) & 2.2 \(\mu\)F \\ \hline
        Current Controller Bandwidth & 3000 Hz \\ \hline
        Voltage Controller Bandwidth & 45 Hz \\ \hline
        Active Balancing Controller Bandwidth & 246 Hz \\ \hline
        Feedback gain (\(\alpha\)) & 0.047 \\ \hline
        Load resistance (\(R_{\text{load}}\)) & 5 \(\Omega\) \\ \hline
    \end{tabular}
    \caption{System Parameters for FCML Control}
    \label{tab:fcml_parameters}
\end{table}

\section{Conclusion}
This article proposes an estimator-based control framework for hybrid FCML converters, addressing limitations of conventional approaches with a hybrid estimation method that combines high-bandwidth closed-loop feedback and rapid open-loop feedforward dynamics. The proposed approach eliminates reliance on isolated voltage sensors by utilizing high-bandwidth flying capacitor voltage estimation. Key contributions include a detailed analysis of stability and gain tuning, and the effects of rank-deficiency. The methodology is proposed to achieve high-bandwidth active voltage balancing and current control with reduced sampling rates, offering a practical and scalable solution for power electronics applications for grid-tied datacenters, electric aircraft, and motor drive.

\ifCLASSOPTIONcaptionsoff
  \newpage
\fi

\bibliographystyle{IEEEtran} 
\bibliography{IEEEabrv}

\end{document}